\documentclass{aa}  
\usepackage{orcidlink}

\usepackage{graphicx}
\usepackage{txfonts}
\usepackage{xcolor}

\hypersetup{
    colorlinks=true,
    linkcolor=blue,
    citecolor=blue,
    urlcolor=blue
}

\usepackage[final]{changes}
\usepackage{natbib}
\bibpunct{(}{)}{;}{a}{}{,} 
\usepackage{academicons}
\definecolor{orcidlogocol}{HTML}{A6CE39}
\newcommand{\orcidicon}[1]{\href{https://orcid.org/#1}{\textcolor{orcidlogocol}{\aiOrcid}}}

\begin{document}

   \title{Radiative cooling effects on plasmoid formation in black hole accretion flows with multiple magnetic loops}

\author{
  Jing-Ze Xia\inst{1}\orcidicon{0009-0004-8669-2411} \and
  Hong-Xuan Jiang\inst{1}\orcidicon{0000-0003-0292-2773} \and
  Yosuke Mizuno\inst{1,2,3,4}\orcidicon{0000-0002-8131-6730} \and
  Antonios Nathanail\inst{5} 
  \and
  Christian M. Fromm\inst{6,4,7}
}

\institute{
  \inst{} Tsung-Dao Lee Institute, Shanghai Jiao Tong University, 1 Lisuo Road, Shanghai 201210, PR China\\
  \email{jackxia@sjtu.edu.cn; hongxuan\_jiang@sjtu.edu.cn; mizuno@sjtu.edu.cn}
  \and
  \inst{} School of Physics and Astronomy, Shanghai Jiao Tong University, 800 Dongchuan Road, Shanghai 200240, PR China
  \and
  \inst{} Key Laboratory for Particle Physics, Astrophysics and Cosmology (MOE), Shanghai Key Laboratory for Particle Physics and Cosmology, Shanghai Jiao Tong University, 800 Dongchuan Road, Shanghai, 200240, PR China
  \and
  \inst{} Institut f\"ur Theoretische Physik, Goethe-Universit\"at Frankfurt, Max-von-Laue-Str 1, D-60438 Frankfurt am Main, Germany
   \and 
   \inst{} Research Center for Astronomy, Academy of Athens, Soranou Efessiou 4, GR-11527 Athens, Greece
   \and
   \inst{} Institut f\"ur Theoretische Physik und Astrophysik, Universit\"at W\"urzburg, Emil-Fischer-Str. 31, D-97074 W\"urzburg, Germany
   \and
   \inst{} Max-Planck-Institut f\"ur Radioastronomie, Auf dem H\"ugel 69, D-53121 Bonn, Germany
}
   \date{Received September 15, 20XX; accepted March 16, 20XX}

  \abstract
   {We investigate the physics of black hole accretion flows, particularly focusing on phenomena like magnetic reconnection and plasmoid formation, which are believed to be responsible for energetic events such as flares observed from astrophysical black holes.}
   {We aim to understand the influence of radiative cooling on plasmoid formation within black hole accretion flows that are threaded by multi-loop magnetic field configurations.}
   {We conducted two- and three-dimensional two-temperature general relativistic magnetohydrodynamic (GRMHD) simulations. By varying the magnetic loop sizes and the mass accretion rate, we explored how radiative cooling alters the accretion dynamics, disk structure, and the properties of reconnection-driven plasmoid chains.}
   {Our results demonstrate that radiative cooling suppresses the transition to the magnetically arrested disk (MAD) state by reducing magnetic flux accumulation near the horizon. It significantly modifies the disk morphology by lowering the electron temperature and compressing the disk, which leads to increased density at the equatorial plane and decreased magnetization. Within the current sheets, radiative cooling triggers layer compression and the collapse of plasmoids, shortening their lifetime and reducing their size, while the frequency of plasmoid events increases. Moreover, we observe enhanced negative energy-at-infinity density in plasmoids near the ergosphere, with its peaks corresponding to plasmoid formation events.}
   {Radiative cooling plays a critical role in shaping both macroscopic accretion flow properties and microscopic reconnection phenomena near black holes. This suggests that radiative cooling may modulate black hole energy extraction through reconnection-driven Penrose processes, highlighting its importance in models of astrophysical black holes.}
    
    \titlerunning{Radiative Cooling in Multi-Loop Accretion Flows}
    \authorrunning{Xia et al.}

   \maketitle
   \nolinenumbers

\section{Introduction} \label{sec:intro}
Over the past few decades, the black hole (BH) candidate at the center of the Milky Way, Sagittarius A$^*$ (hereafter Sgr~A$^*$), has served as an exceptional laboratory for studying the physics of accretion and outflows due to its proximity to Earth. Numerous observational and theoretical studies have been conducted in-depth on the morphology, variability, and other characteristics of Sgr~A$^*$ \citep{2022AAS...24021101M,2022ApJ...930L..12E,2022ApJ...924L..32R}. Episodic flares are detected across the electromagnetic spectrum, attracting great interest from the astronomical community.

Despite extensive efforts, the precise mechanism responsible for these flares remains under debate. Several mechanisms have been proposed, including flux eruption events \citep{2021MNRAS.502.2023P,2025A&A...696A..10A}, magnetic reconnection \citep{2024MNRAS.531.3136L,2023MNRAS.520.1271L}, and plasmoid chains along current sheet \citep{2025A&A...696A..36D}.
These processes collectively contribute to the observed variability and energetic phenomena near the BH \citep{2007MNRAS.375..764T,2004ApJ...606..894Y}. 
Observations from the Event Horizon Telescope (EHT) on Sgr~A$^*$ suggest the presence of a strong and ordered magnetic field \citep{2024ApJ...964L..25E}. These results indicate a preference for the magnetically arrested disk (MAD) regime compared to the Standard and Normal Evolution (SANE) regime and other types of accretion flows \citep{2022ApJ...930L..12E,2024ApJ...964L..25E,2022ApJ...930L..16E}. The absence of a powerful jet in current observations, typically associated with the MAD accretion regime \citep{2024ApJ...974..116C, 2022MNRAS.511.3795N}, suggests the incompleteness of our current models.

Large-scale magnetized jets generated by accretion flows around BHs often exhibit variable energy output \citep{2020MNRAS.495.1549N}. 
Due to the inherently turbulent nature of the accretion process \citep{2022ApJ...935L...1L}, magnetic field loops of different polarities are generated around the BH. The merging of the magnetic loops causes frequent polarity inversion events and forms current sheets \citep{2022ApJ...924L..32R,2023A&A...677A..67E,2015PhPl...22a2902K}. The tearing instability in the current sheets creates plasmoids and x-points, making them ideal places for particle acceleration and the flaring events of Sgr~A$^*$ via magnetic reconnection\citep{2009ApJ...699.1789T,2017MNRAS.467.3279R,2019MNRAS.485..299R,2018A&A...618L..10G,2023A&A...677L..10G,2022ApJ...924L..32R}. 

High-resolution GRMHD simulations show that mean-field dynamo and magnetorotational instability can transform toroidal magnetic fields into poloidal loops, creating a multi-loop magnetic configuration in accretion tori \citep{2020MNRAS.494.3656L, 2023ApJ...954...40K, 2023ApJ...954L..21G, 2024ApJ...960...97R}. To mimic this process, simulations with initial multiloop magnetic field configuration have been performed to study the dynamics of the accretion flows \citep{2020MNRAS.495.1549N,2022MNRAS.513.4267N,2023MNRAS.522.2307J,2024A&A...688A..82J}. The alternating polarity configuration not only prevents the accumulation of magnetic fields but also facilitates frequent reconnection between magnetic loops of opposite polarities, thereby releasing magnetic energy. A series of two-dimensional (2D) GRMHD simulations with multi-loop configurations was conducted by \cite{2020MNRAS.495.1549N}, while three-dimensional (3D) simulations with multi-loop magnetic field configurations were performed by \cite{2021MNRAS.508.1241C} and \cite{2022MNRAS.513.4267N}. These studies demonstrate that the multi-loop magnetic field configuration can efficiently produce energetic plasmoids via magnetic reconnection, which have been proposed to explain flaring events from Sgr~A$^*$ \citep{2016MNRAS.462.3325P,2018ApJ...862L..25Z,2019MNRAS.482...65C}. Another possible explanation for these flares is the energy release during flux eruption events in MAD state, which also provide suitable conditions for accelerating non-thermal particles \citep{2021MNRAS.508.1241C,2022MNRAS.511.3536S,2023MNRAS.522.2307J,2024MNRAS.530.1563G}. Unlike SANE models, which typically produce only weak jets, MAD models generally launch powerful jets. However, strong observational evidence for such a powerful jet in Sgr A$^*$ remains unclear \citep{2022ApJ...930L..12E}. 

Recent studies suggest that radiative cooling can reduce jet efficiency by a factor of approximately 2 \citep{2025ApJ...981L..11S}, making the MAD configuration with radiative cooling a more plausible explanation for Sgr~A$^*$.
The bolometric luminosity of Sgr~A$^*$ is extremely low, $L_{\text{bol}} \sim 10^{36} \, \text{erg s}^{-1} \approx 10^{-9} L_{\text{Edd}}$, where $L_{\text{Edd}}$ represents the Eddington luminosity \citep{2022ApJ...930L..12E}.  Given this extremely low luminosity, radiative cooling has conventionally been regarded as negligible, as it is unlikely to substantially affect the dynamics of the accretion flow. Most previous studies employing 3D GRMHD simulations have omitted radiative cooling losses for simplicity \citep{2018ApJ...864..126R,2023MNRAS.522.2307J,2024A&A...688A..82J}. While this assumption is generally reasonable, \cite{2012MNRAS.426.1928D} argued, based on their 2D simulations, cooling losses become increasingly important at higher accretion rates, i.e. $\dot{M} \gtrsim 10^{-7}\dot{M}_{\text{Edd}}$. Similar results were also obtained by \cite{2023MNRAS.518..405D}.
These losses may influence the dynamics and spectra of Sgr~A$^*$, even within the permitted range of accretion rates inferred from polarization and X-ray studies\citep{2025MNRAS.537.2496C,2025MNRAS.538..698S}.
It is worth noting that single-fluid GRMHD simulations incorporating radiative cooling have also been employed to investigate hot accretion flows around Sgr~A$^*$ \citep{2009ApJ...693..771F,2012MNRAS.426.1928D,2013MNRAS.431.2872D,2020MNRAS.499.3178Y}. The influence of radiative cooling on accretion flow dynamics intensifies with the accretion rate. Specifically, when the accretion rate exceeds approximately $10^{-7} \dot{M}_{\rm Edd}$, radiative cooling plays an increasingly significant role in shaping the dynamical properties of the accretion disk \citep{2020MNRAS.499.3178Y} and their horizon-scale images (Zhang et al. 2025).

With the advancement of astronomical observations and numerical simulations, BHs have emerged as unique laboratories for testing plasma electrodynamics under extreme conditions \citep{2014PhRvD..89b4041L}. Among the most profound predictions of general relativity is the possibility of extracting rotational energy from a BH via the inflow of negative energy and angular momentum across the event horizon \citep{1971NPhS..229..177P}. Recently, the nonlinear nature of magnetic reconnection and plasma dynamics near the surface of the accretion tori around BHs has been increasingly recognized \citep{2015PhRvL.114k5003A}, and it has been reported that energy extraction can occur in the formation of plasmoid chains via Penrose process (PP) \citep{2024PhRvD.110j4044F,2025PhRvD.111b3003S}. Most of these studies have been conducted within purely analytical frameworks. In this work, we extend the investigation by performing numerical simulations and analysis within a multi-loop configuration that incorporates the effects of radiative cooling.

In this study, we perform a series of 2D two-temperature GRMHD simulations with a multi-loop initial magnetic field configuration to investigate the effect of radiative cooling. Specifically, we study the influence of radiative cooling on plasmoid chains, conducting statistical analysis on their properties, including radii and occurrence rates. 
To confirm the generality of the conclusions obtained from 2D simulations, we also perform 3D GRMHD simulations and high resolution 2D GRMHD simulations to further study the disk and temperature properties caused by radiative cooling.

This paper is organized as follows. In Sect.~\ref{sec:style}, we describe the simulation setup. In Sect.~\ref{sec:dynamics}, we report our main results on the effects of radiative cooling. Finally, Sect.~\ref{conclusion} summarizes our main findings and gives some discussion.

\section{Numerical setup} \label{sec:style}

Our GRMHD simulations are conducted with the {\tt BHAC} code \citep{2017ComAC...4....1P,2019A&A...629A..61O}.
{\tt BHAC} code solves the ideal MHD equations in geometric units ($G=M = c = 1$), which are given in equations expressed in covariant notation as follows:
\begin{equation}
\begin{aligned}
\nabla_\mu (\rho u^\mu) &= 0, \\
\nabla_\mu T^{\mu\nu} &= 0, \\
\nabla_\mu {}^*F^{\mu\nu} &= 0,
\end{aligned}
\label{Eq: GRMHD}
\end{equation}
which express the conservation of mass, the energy-momentum tensor $T^{\mu\nu}$, and the homogeneous Maxwell equations, respectively. The Faraday tensor $F^{\mu\nu}$ further characterizes the electromagnetic fields in the system. In GRMHD codes, Eq.~\ref{Eq: GRMHD} is transformed into conservative form through 3+1 formalism which is written by \citep{2007A&A...473...11D, 2017ComAC...4....1P}:
\begin{eqnarray}
\partial_t (\sqrt{\gamma} \mathbf{U}) + \partial_i (\sqrt{\gamma} \mathbf{F}^i) = \sqrt{\gamma} \mathbf{S},
\label{eq2}
\end{eqnarray}
where the specific formulation of terms $\mathbf{U}$, $\mathbf{F}$ and $\mathbf{S}$ are the conserved variables, fluxes and source term. Following \citet{2023MNRAS.518..405D}, we incorporate radiative cooling by implementing the modified source term $S = S_0 + S^{\prime}$, where $S^{\prime}$ accounts for contributions from radiative cooling processes, as detailed in \citet{2023MNRAS.518..405D}.
The explicit form of the additional source term $S^{\prime}$ is expressed as:
\begin{equation}
\begin{aligned}
    S^{\prime} &= 
    \begin{bmatrix}
        0 \\
        -\alpha \gamma v_j \Lambda \\
        -\alpha \gamma \Lambda \\
        0
    \end{bmatrix}
\end{aligned}
\end{equation}
where $\alpha$ is the lapse function, $\gamma$ is the Lorentz factor, $v_j$ represents the fluid three-velocity, and $\Lambda$ is the total radiative cooling term, which includes synchrotron cooling and bremsstrahlung cooling.
The explicit form of dissipative heating, Coulomb interaction, and radiative cooling processes are introduced in Sect.~\ref{cooling}.

\subsection{Model setup}\label{setup}

The simulations are initialized with Fishbone-Moncrief hydrodynamic equilibrium torus \citep{1976ApJ...207..962F}, with parameters $r_{\text{in}} = 20 \, r_{\rm g}$ and $r_{\text{max}} = 40 \,r_{\rm g}$, where $r_{\rm g} \equiv GM / c^2$ is the gravitational radius, $G$ is gravitational constant and $M$ is the BH mass. An ideal gas equation of state is adopted, with a constant adiabatic index of $\Gamma=4/3$.

In all cases, we consider a purely poloidal magnetic field as an initial condition. The initial magnetic configuration is given by defining the vector potential as follows:
\begin{eqnarray}
A_\phi \propto (\rho - 0.01) \left(\frac{r}{r_{\text{in}}}\right)^3 \sin^3\theta \exp\left(-\frac{r}{400}\right).
\end{eqnarray}
To control the number of magnetic loops in the vertical and radial directions, the vector potential $A_\phi$ is further multiplied by $\cos((N - 1)\theta)$ and $\sin\left(2\pi (r - r_{\text{in}})/\lambda_r\right)$, where $\lambda_r$ represents the radial loop wavelength, and $N$ is the number of magnetic loops in the $\theta$-direction. For all cases, following \cite{2023MNRAS.522.2307J}, we set $N$ = 3.

Our simulations use spherical modified Kerr–Schild coordinates. The simulation domain extends radially from the BH event horizon to $r = 2500 \, r_{\rm g}$ with a logarithmic spacing. In the polar direction, the domain spans from the polar angle $\theta$ = 0 to $\pi$ with uniform spacing. In all 2D simulations, we adopt a standard resolution of $1024 \times 512$, except for one high-resolution case with an effective resolution of $4096 \times 2048$ (see Appendix~\ref{appenD} for details on the numerical setup). For the 3D simulations, we employ two levels of static mesh refinement, achieving an effective resolution of $320 \times 160 \times 160$. The 2D simulations are evolved until $t = 15\,000 \,\rm  M$, while the 3D simulations are extended to $t = 25\,000 \,\rm  M$ to reach quasi-steady state.

To excite the magnetorotational instability (MRI), we apply a $2\%$ random perturbation to the gas pressure within the torus. We adopt a floor treatment to ensure numerical stability in low-density regions, particularly near the BH and the rotation axis 
 \citep{2015GReGr..47....3G}. Specifically, the floor values for the rest-mass density and pressure are set to $\rho_{\text{fl}} = 10^{-4} r^{-3/2}$ and $p_{\text{fl}} = (10^{-6}/3) r^{-5/2}$, respectively. For the electron pressure, we use $p_e = 0.01 p_{\text{fl}}$ for $p_e \leq 0.01 p_{\text{fl}}$ and $p_{\rm e} = 0.99 p_{\rm g}$ for $p_{\rm e} \geq 0.99 p_{\rm g}$, where $p_{\rm g}$ is gas pressure. Additionally, the entropy of the gas ($\kappa_{\rm g}$) and the electrons ($\kappa_{\rm e}$) are recalculated, where the floor condition is applied.
In all simulations, the BH mass and spin are set to $M_{\rm BH} = 4.5 \times 10^6 \,\rm M_\odot$ and $a_\ast = 0.9375$, respectively.

To conduct a comprehensive investigation of various scenarios, we consider several simulation models. The details of these models are summarized in Table \ref{tab:model_parameters}.

\subsection{Radiative cooling}\label{cooling}

Following \citet{2015MNRAS.454.1848R} and \citet{2017MNRAS.466..705S}, we solve the electron-entropy equation in the presence of dissipative heating, Coulomb interaction, and radiative cooling processes:
\begin{eqnarray}
T_e \nabla_\mu (\rho u^\mu \kappa_e) = f_{\rm e} Q + \Lambda_{\text{ei}} - \Lambda,
\label{eq5}
\end{eqnarray}
where $\kappa_{\rm e} := \exp[(\Gamma_{\rm e} - 1) s_{\rm e}]$ and $s_{\rm e} := p / \rho^{\Gamma_{\rm e}}$ represent the electron entropy per particle. Here, $\Gamma_{\rm e}$ denotes the adiabatic index of the electrons where we choose $\Gamma_{\rm e}=4/3$, $Q$ is the rate of dissipative heating, and $f_{\rm e}$ is the fraction of dissipative heating transferred to the electrons. $\Lambda_{\text{ei}}$ is the energy transferred to the electrons from the ions through Coulomb interaction, and $\Lambda$ represents the radiative cooling term. 
The radiative cooling rate is computed following \citet{2012MNRAS.426.1928D,2023MNRAS.518..405D}, using the solutions from \citet{1996ApJ...465..312E} to account for bremsstrahlung, synchrotron, and inverse Compton cooling processes (see Appendix~\ref{appenC} for detailed calculations).
Solving the fluid equations (Eq.~\ref{Eq: GRMHD}) together with the electron-entropy equation (Eq.~\ref{eq5}) yields all relevant properties of the accretion flows and electron thermodynamics.

\begin{table}
\caption{Model parameters including $\lambda_r$, mass accretion rate normalized by Eddington ratio, radiative cooling status, and grid number.}
\label{tab:model_parameters}
\centering
\begin{tabular}{ccccc}
\hline\hline
Model & $\lambda_r$ & $ \dot{M}/\dot{M}_{\text{Edd}}$ & Radiative cooling & Resolution \\
\hline
A & 40 & $1.0 \times 10^{-3}$ & Yes & $1024\times512$  \\
B & 40 & $1.0 \times 10^{-5}$ & Yes & $1024\times512$  \\
C & 40 & - & No & $1024\times512$  \\
D & 80 & $1.0 \times 10^{-3}$ & Yes & $1024\times512$  \\
E & 80 & $1.0 \times 10^{-5}$ & Yes & $1024\times512$  \\
F & 80 & - & No & $1024\times512$  \\
\tt{M40a3d} & 40 & $1.0 \times 10^{-3}$ & Yes & $320\times160\times160$  \\
\tt{M40n3d} & 40 & - & No & $320\times160\times160$  \\
\tt{hr} & 40 & $1.0 \times 10^{-3}$ & Yes & $4096\times2048$  \\
\hline
\end{tabular}
\end{table}

\section{Influence of radiative cooling on accretion flows} \label{sec:dynamics}

\subsection{Comparison of properties in the accretion flow} \label{subsec:accretion}

This section investigates the impact of radiative cooling on the accretion process by analyzing the mass accretion rate $\dot{M}$ and magnetic flux $\Phi_{\rm B}$ through the BH horizon. Our 2D simulations include six different runs with different initial magnetic field configurations and radiative cooling setup.
Following \citet{2019ApJS..243...26P}, we define the mass accretion rate $\dot{M}$ and magnetic flux $\Phi_{\rm BH}$ near the BH horizon as follows:
\begin{equation}
\dot{M} = \int_0^{2\pi} \int_0^{\pi} \rho u^r \sqrt{-g} \, d\theta \, d\phi,
\end{equation}
\begin{equation}
\Phi_{\rm BH} = \frac{1}{2} \int_0^{2\pi} \int_0^{\pi} |B^r| \sqrt{-g} \, d\theta \, d\phi,
\end{equation}
where $\rho$ is the rest-mass density, $u^r$ is the radial component of the four-velocity, $B^r$ is the radial magnetic field, and $\sqrt{-g}$ is the square root of the metric determinant.

\begin{figure*}
    \centering
    \includegraphics[width=\linewidth]{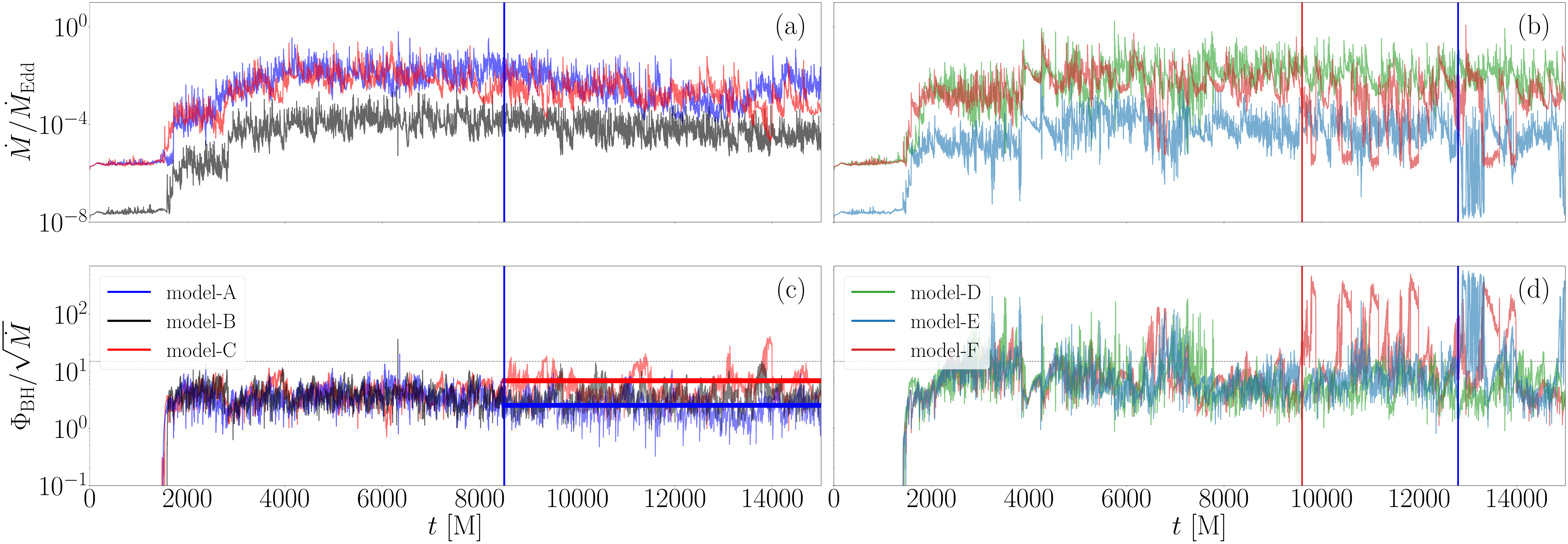}
    \caption{Time evolution of mass accretion rates measured at the event horizon (top) and normalized magnetic flux at the horizon (bottom). The left panels (a,c) correspond to the case with smaller magnetic loops ($\lambda = 40$), while the right panels (b,d) represent the case with larger magnetic loops ($\lambda = 80$). 
    The curves in different colors correspond to the different radiative cooling and time-averaged accretion rates: radiative cooling with $\dot{M}/\dot{M}_{\rm Edd}=1 \times10^{-3}$ (bule, green), radiative cooling with $1 \times10^{-5}$(black, light-blue), and no cooling (red, light-red).
    In panel (c), the horizontal red and blue lines denote the average magnetic flux of models A and C, respectively.
    The blue vertical lines in panels (a) and (c) mark the onset of the quasi-steady state. In panels (b) and (d), the light-blue and light-red vertical lines indicate the transition from SANE to MAD for models {\tt E} and {\tt F}, respectively.
    The time is given in units of the light crossing time, $t_g \equiv GM/c^3 = [\text{M}]$.
    } 
    \label{fig:Fig1}
\end{figure*}

In Fig.~\ref{fig:Fig1}, the time evolutions of mass accretion rate $\dot{M}$ and normalized magnetic flux $\Phi_{\rm BH}/\sqrt{\dot{M}}$ of the models are presented. Panels (a) and (c) of Fig.~\ref{fig:Fig1} depict models initialized with smaller magnetic loops (wavelength $\lambda_r = 40$), while panels (b) and (d) show models with larger magnetic loops (wavelength $\lambda_r = 80$). Different colors in these panels indicate variations in the values of mass accretion rate normalized by Eddington ratio, $\dot{M}_{\rm Edd}$, and with/without radiative cooling. The dashed lines in the panels represent $ \Phi_{\text{BH}} / \sqrt{\dot{M}} = 15$, which is the criterion for MAD regime \citep{2011MNRAS.418L..79T,2021MNRAS.506..741M}.

From the magnetic flux evolutions in Fig.~\ref{fig:Fig1}, we observe a decreased magnetic flux from the models with stronger cooling. The vertical blue lines in the panel indicates the moment the accretion flow achieves a quasi-steady state for smaller magnetic loop cases. The horizontal lines represent the average normalized magnetic flux during this period. In the quasi-steady state, the non-cooling case (red line) exhibits greater magnetic flux accumulation compared to the case with radiative cooling (blue line). 
Due to the relatively smaller magnetic loops, the accretion flows do not reach the MAD regime \citep{2023MNRAS.522.2307J}. Radiative cooling mostly reduces the magnetic flux, which is also evident in models D, E, and F in the larger loop runs in Fig.~\ref{fig:Fig1}(d).
We note, however, that the cooling-induced reduction of the normalized magnetic flux at the event horizon may not be universal. Recent work by \cite{2025ApJ...981L..11S} demonstrated that the response of the MAD parameter to radiative cooling can depend sensitively on the black hole spin, with different spin values exhibiting distinct trends in magnetic flux accumulation. In the present study, we restrict our analysis to a single black hole spin configuration and therefore do not explore this additional parameter space. We focus on a rapidly spinning black hole with $a = 0.9375$, motivated by observational constraints suggesting that Sgr~A$^*$ likely possesses a high spin \citep{2022ApJ...930L..16E}. Consequently, while our results indicate a systematic suppression of magnetic flux with increasing cooling strength for the spin considered here, a broader survey over black hole spins will be required to establish the generality of this behavior.

\begin{figure}
    \centering
    \includegraphics[width=\linewidth]{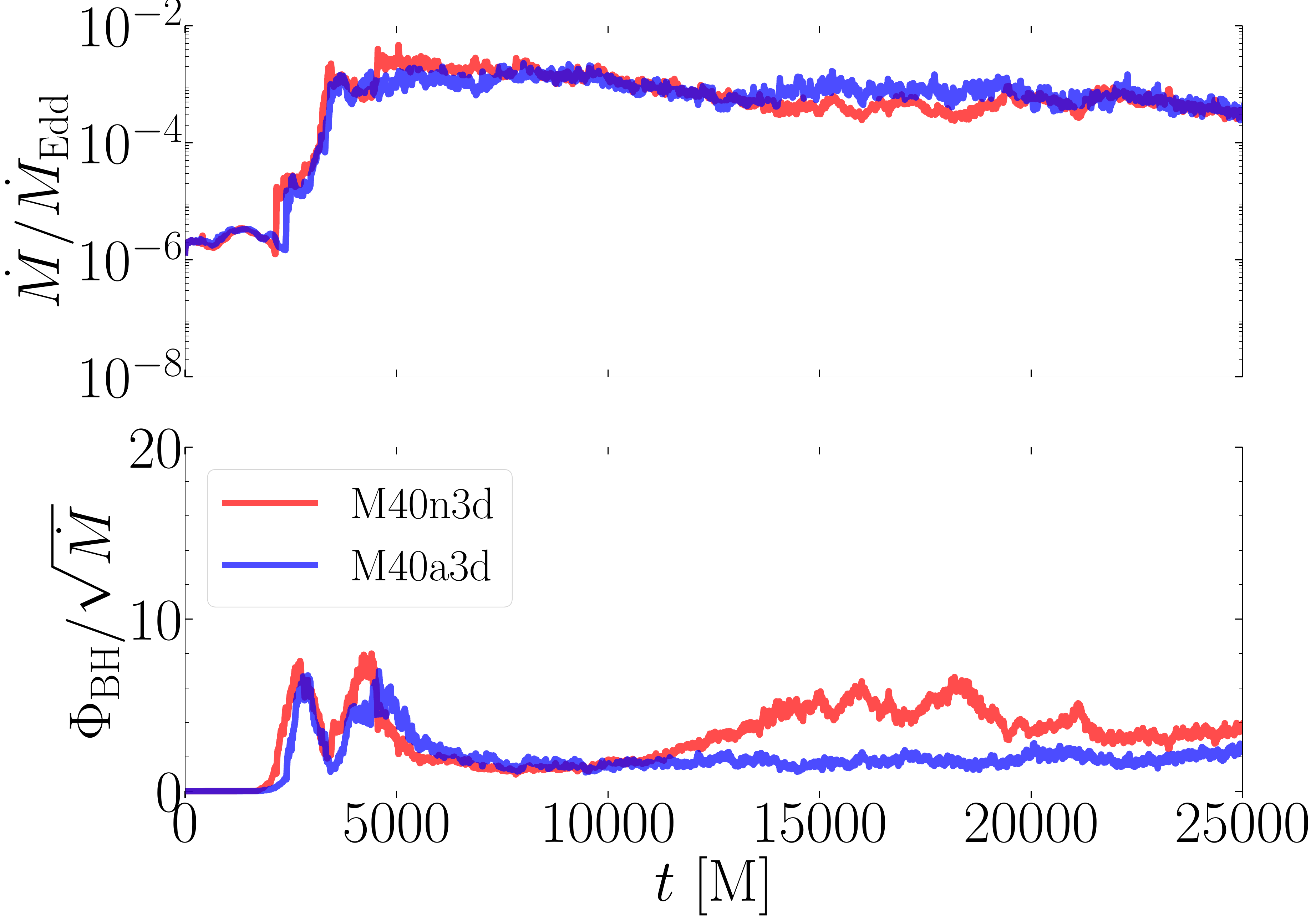}
    \caption{Same as Fig.~\ref{fig:Fig1} but shown in 3D cases (\tt{M40n3d} and \tt{M40a3d}).
    }
    \label{fig:Mdot_3D}
\end{figure}

For the larger loop models (wavelength $\lambda_r = 80\,r_{\rm g}$), shown in the right panels of Fig.~\ref{fig:Fig1}, fewer magnetic loops within the torus result in reduced magnetic dissipation between them. The merging of these loops generates abundant small-scale magnetic fields, which initially prevent significant magnetic field accumulation near the BH, producing a SANE-type flow in the early stages of the simulations. As these small-scale fields coalesce into a more ordered, large-scale magnetic structure, the accretion flow transitions from a SANE to a MAD regime, as observed in Panel (d) of Fig.~\ref{fig:Fig1} after the vertical lines.
The light-blue and light-red vertical lines indicate the transition time from SANE to MAD regime. From model E (light-blue lines), which has a relatively weak cooling effect, we see a delayed transition to the MAD regime compared to the non-cooling case (model F, light-red lines). While in model D (green lines), the cooling effect is so strong that the SANE to MAD transition does not occur within the simulation duration.
Radiative cooling reduces the accumulation of magnetic flux near the BH horizon, resulting in insufficient magnetic flux pile-up to reach the MAD regime. 

Similar behavior is also seen in the 3D simulations in Fig.~\ref{fig:Mdot_3D}. Radiative cooling suppresses the accumulation of magnetic flux, consistent with the conclusions drawn from our 2D simulations.

\subsection{The influence of radiative cooling} \label{subsec:influence}

\subsubsection{Electron temperature} \label{temperature}
\begin{figure*}
\centering
\includegraphics[width=12cm]{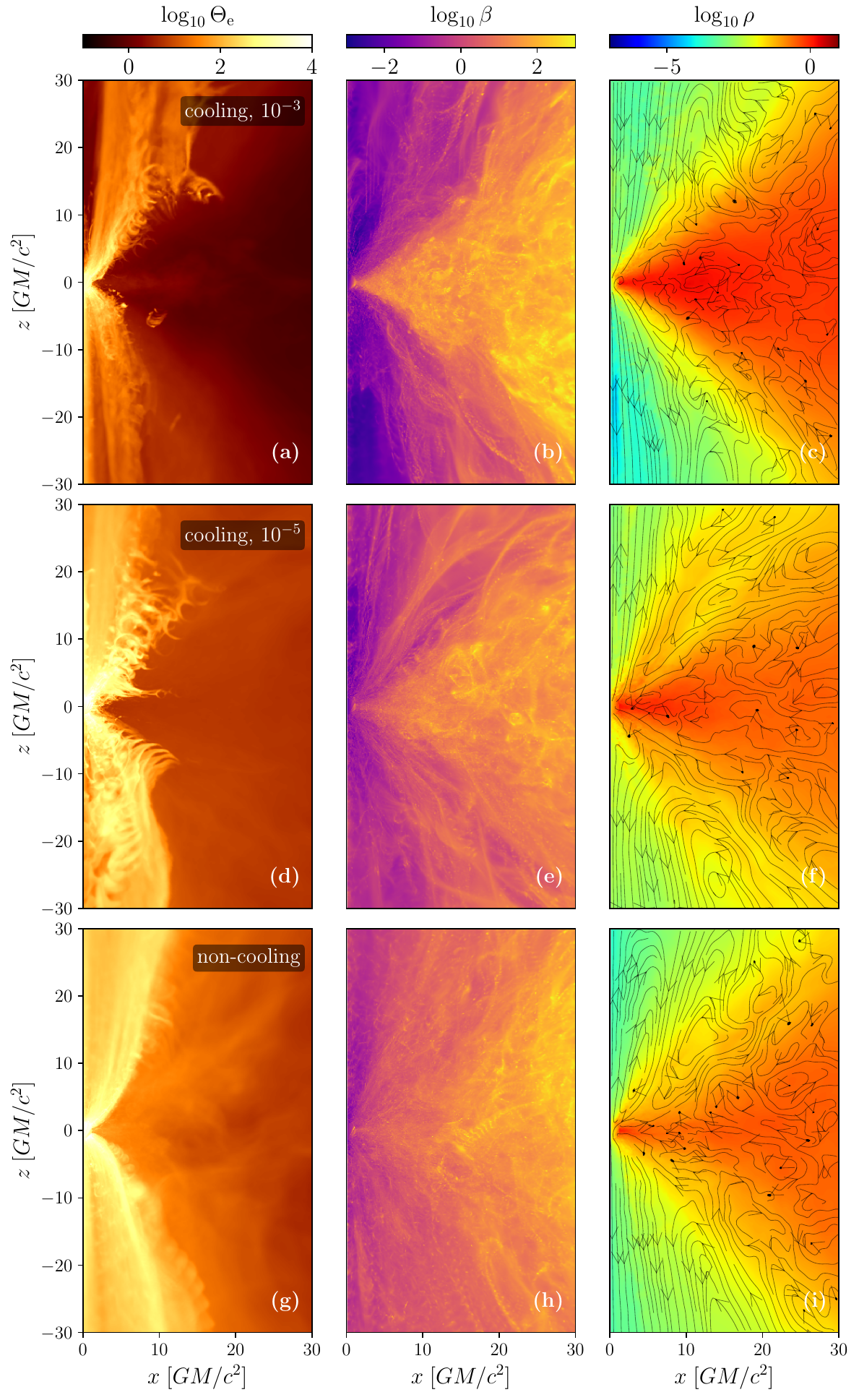}
\caption{Time-averaged electron temperature $\Theta_{\rm e}$, plasma $\beta$, and density $\rho$ distributions for three models. From top to bottom, the panels correspond to models A, B, and C. The plasma beta parameter is defined as $\beta=p_{\rm g}/p_{\rm mag}$, where $p_{\rm g}$ represents the gas pressure, and $p_{\rm mag}=b^2/2$ denotes the magnetic pressure. The light black contours in the density profile represent the magnetic field. The average is taken over the time range $t=8000\,{\rm M}$ to $13000\,{\rm M}$.}
\label{fig:Fig3}
\end{figure*}

The most direct impact of radiative cooling on our simulations is the substantial alteration of electron temperature. 
To illustrate the effects of radiative cooling on the physical structure of the disk, in Fig.~\ref{fig:Fig3}, we show the time-averaged profiles of the electron temperature $\Theta_{\rm e}$, plasma beta $\beta$, and density $\rho$ for Models A, B, and C. These quantities are averaged over the time interval $t =$ 8000 -- 13000$\,\mathrm{M}$, during which the mass accretion rate remains approximately steady.
We use the two-temperature approach to calculate the time evolution of electron entropy following \citep{2015MNRAS.454.1848R,2021MNRAS.506..741M}. The energy injection ratio to electron and ion comes from the prescription based on turbulent plasma in \cite{2019PNAS..116..771K}, see more details in \cite{2021MNRAS.506..741M, 2023MNRAS.522.2307J}. 
The electron pressure $p_{\,\rm e}$ is calculated as $p_{\,\rm e} = \kappa_{\,\rm e} \rho^{\Gamma_{\rm e}}$, where the value of $\kappa_{\,\rm e}$ is updated according to Eq.~3 in \cite{2021MNRAS.506..741M}. The ion-to-electron temperature ratio is then determined as $T_{\,\rm i} / T_{\,\rm e} = (p_{\,\rm g} - p_{\,\rm e}) / p_{\,\rm e}$. The dimensionless electron temperature $\Theta_{\,\rm e}$ is defined as:
\begin{eqnarray}
    \Theta_{\,\rm e} = \left( \frac{p_{\,\rm g} - p_{\,\rm e}}{\rho} \right) \left( \frac{m_{\,\rm p} / m_{\,\rm e}}{T_{\,\rm i} / T_{\,\rm e}} \right),
\end{eqnarray}
where $p_{\,\rm g}$ is the total gas pressure, $\rho$ is the density, $m_{\,\rm p}$ and $m_{\,\rm e}$ are the proton and electron masses, respectively, and $T_{\,\rm i} $ and $T_{\,\rm e}$ are the ion and electron temperatures.

\begin{figure}
    \centering
    \includegraphics[width=\linewidth]{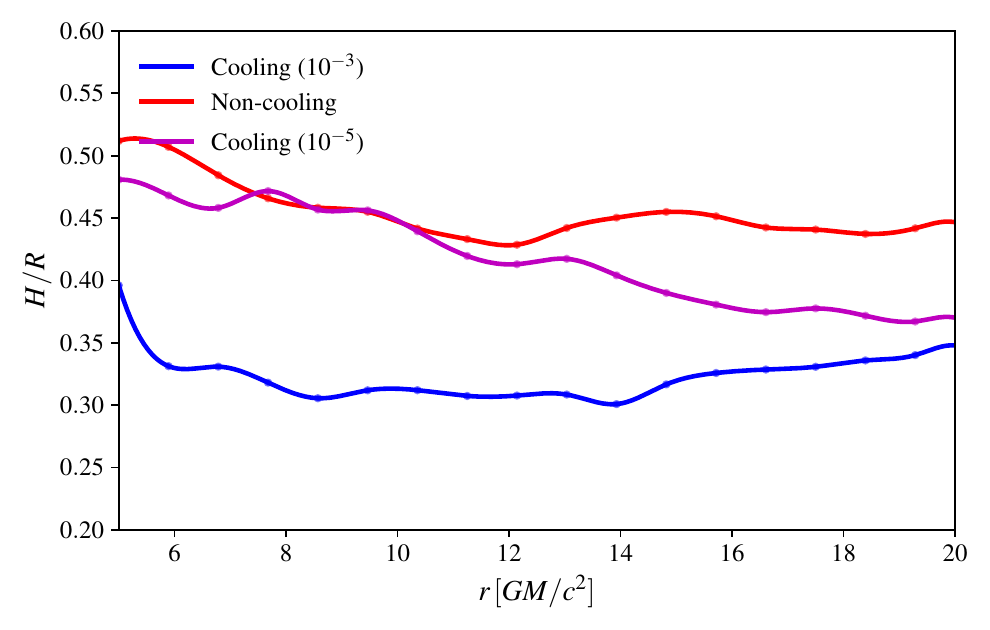}
    \caption{
The density-weighted scale height as a function of radius for various cooling models A, B, and C.
    }
    \label{fig:H/r}
\end{figure}

In the left column of Fig.~\ref{fig:Fig3}, the electron temperature of the simulations with different magnitudes of cooling are presented. We confirm that radiative cooling reduces the overall electron temperature, particularly in the disk region. The extent of this reduction depends on the cooling intensity, consistent with the findings of \citet{2023MNRAS.518..405D}.
However, radiative cooling not only globally decreases the electron temperature but also further impacts the associated dynamics and structure of the accretion disk.
Due to the radiative loss of thermal energy, the gas pressure is reduced, leading to a lower scale height of the disk. As shown in the right column of Fig.~\ref{fig:Fig3}, we present the average density distributions of models A, B, and C. It is evident that the temperature reduction induced by radiative cooling significantly increases the density in the disk, resulting in vertical compression in disk height.
To quantify the geometric structure of the disk, we compute the density-weighted scale height \(H\), following the definition in \citet{2009ApJ...692..411N}:
\[
H \equiv \left( \langle z^2 \rangle - \langle z \rangle^2 \right)^{1/2},
\]
where \(z = r \cos \theta\), and the angle brackets \(\langle \cdot \rangle\) denote density-weighted shell averages:
\[
\langle f \rangle = \frac{\int_{\sigma<1} f \rho \sqrt{-g} \, d\theta d\phi}{\int_{\sigma<1} \rho \sqrt{-g} \, d\theta d\phi}.
\]
Here, \(\sigma < 1\) denotes the integration domain within the disk region. As shown in Fig.~\ref{fig:H/r}, radiative cooling leads to a clear reduction in the disk scale height, particularly in the inner regions.
This flattening of the disk reflects the loss of internal energy and thermal pressure support due to radiative losses. For models with stronger radiative cooling, the increase in density compared to the no-cooling model is more pronounced and extends over a wider radial extent. 

Furthermore, the middle column of Fig.~\ref{fig:Fig3} illustrates that radiative cooling leads to the formation of a more ordered and high-beta region, which in turn suppresses the effectiveness of the MRI in shaping the disk structure and contributes to a thinner disk. 

\begin{figure}
    \centering
    \includegraphics[width=\linewidth]{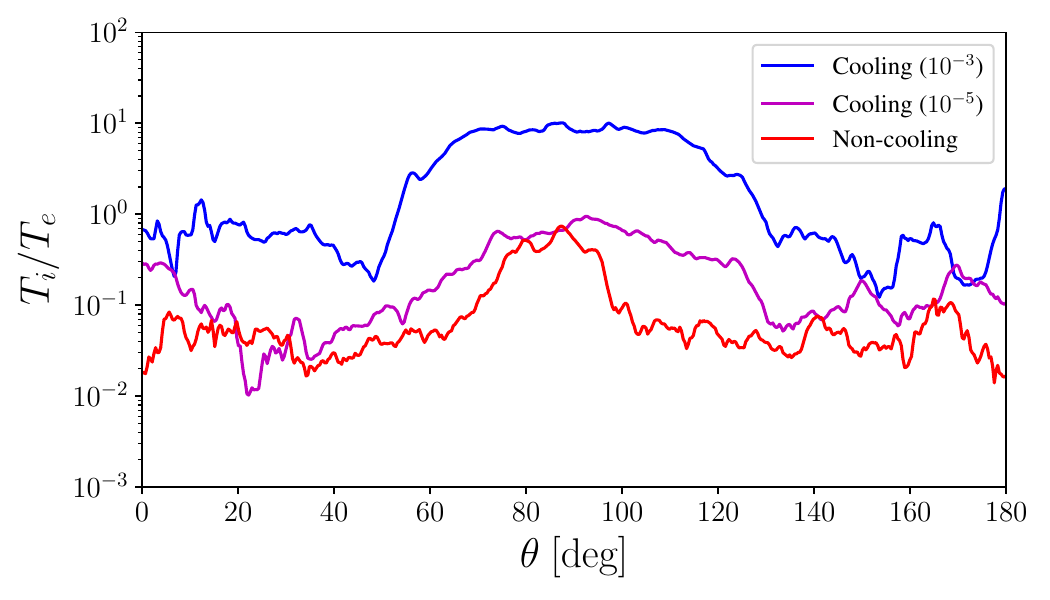}
    \caption{Azimuthal and time-averaged polar-angle profiles of the ion-to-electron temperature ratio of the models A, B, and C at $r=15\, \rm M$. Using the same averaging time range as in Fig.~\ref{fig:Fig3}.}
    \label{fig:TiTe}
\end{figure}

Fig.~\ref{fig:TiTe} presents azimuthal- and time-averaged polar-angle profiles of the ion-to-electron temperature ratio of different cooling models at $r=15\, \rm M$. We observe that both model A (with cooling) and model C (without cooling) exhibit a higher ion-to-electron temperature ratio near the equatorial plane compared to other polar angles; similar plots have been shown by \cite{2021MNRAS.506..741M}. Model A tends to exhibit a globally higher ion-to-electron temperature ratio compared to both the no-cooling case (model C) and the weaker cooling case (model B). It is worth noting that in both model B and model C, the ion temperature remains lower than the electron temperature (\(T_{\,\rm i} < T_{\,\rm e}\)) throughout the domain. In contrast, model A exhibits a region near the equatorial plane where \(T_{\,\rm i} > T_{\,\rm e}\), due to the radiative cooling, which primarily reduces the electron temperature in the disk region. While model A still shows \(T_{\,\rm i} < T_{\,\rm e}\) toward the polar regions, consistent with the other models. This behavior arises because most of the electron heating occurs around the funnel region. 
Consistent with our expectations, the high-resolution simulations reproduce similar trends observed in the low-resolution runs, confirming the influence of radiative cooling on the thermodynamic structure of the accretion flows.

\begin{figure}
    \centering
    \includegraphics[width=0.9\linewidth]{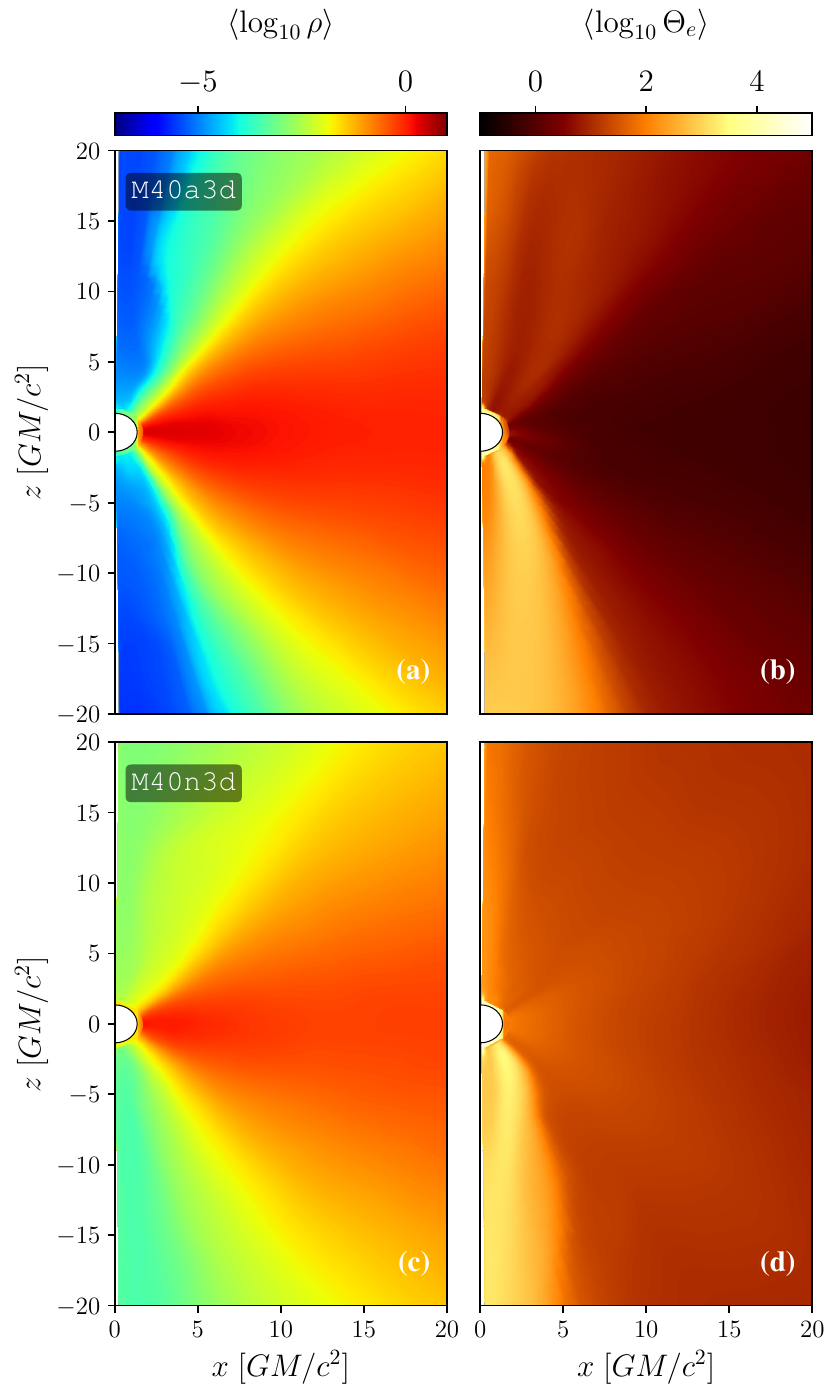}
    \caption{Distribution of the time-averaged density and electron temperature at the $\phi = 0$ for the models (with cooling and without cooling), averaging range $t = 10000\, \rm M$--$14000\,\rm M$. 
}
    \label{fig:3D-rho}
\end{figure}

In our 3D simulations, the disk exhibits similar features under the influence of radiative cooling. As shown in Fig.~\ref{fig:3D-rho}, radiative cooling leads to a noticeable vertical compression of the disk, which in turn results in an increase in the density.
Interestingly, in model \texttt{M40a3d} shown in Fig.~\ref{fig:3D-rho}, we clearly observe that radiative cooling leads to a lower density in the funnel region. Radiative cooling also causes more matter to accumulate near the equatorial plane, resulting in a cleaner (i.e., lower-density) environment away from the midplane. Moreover, reducing gas pressure in the disk suppresses the radial transport of mass and magnetic flux to the BH. Because the magnetic flux threading the black hole is reduced, it is limiting the power available for the Blandford--Znajek mechanism. Thus, jet efficiency becomes decreased with radiative cooling \citep{2025ApJ...981L..11S}.

Consistent with the findings in the 2D simulations of \cite{2023MNRAS.522.2307J}, we observe that a smaller magnetic loop configuration can lead to the emergence of a one-sided jet. The same conclusions are obtained across simulations with different dimensionality. 

\begin{figure}
    \centering
    \includegraphics[width=\linewidth]{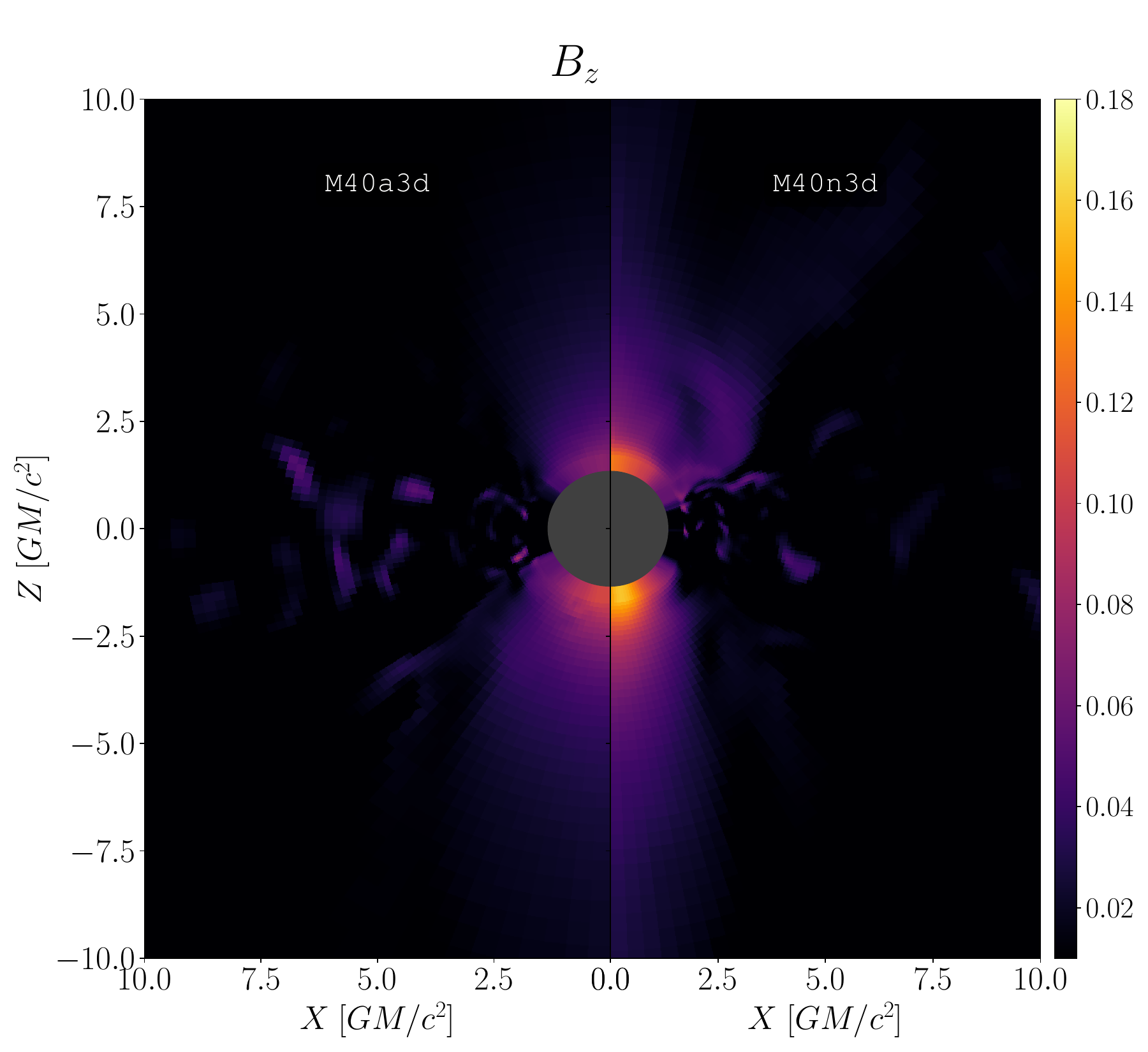}
    \caption{The distribution of axial component of magnetic field $B_{z}$ in a cross-sectional slice at $t = 7920\,\rm M$, with the left and right panels corresponding to models {\tt M40a3d} and {\tt M40n3d}, respectively. 
   }
    \label{fig:3d-B_phi}
\end{figure}

As discussed in Sect.~\ref{subsec:accretion}, radiative cooling also reduces the accumulation of poloidal magnetic flux. This effect is likewise captured in the cross-sectional slices of our 3D simulations as shown in Fig.~\ref{fig:3d-B_phi}. A more detailed discussion of this phenomenon is provided in Sect.~\ref{flux}.

\subsubsection{Radiative collapse and plasmoid formation} \label{collaspe}

Besides of the global effects of radiative cooling on the accretion flow properties, it also significantly affects small-scale structures.
When radiative cooling is included, the reconnection layer exhibits noticeable instabilities. Plasmoids formed via tearing instability tend to collapse due to enhanced cooling effects. This process, known as radiative collapse, occurs when cooling-driven compression further amplifies the cooling rate, leading to a so-called runaway collapse of the current sheet \citep{1995ApJ...449..777D, 2011PhPl...18d2105U}.
As radiative losses deplete the internal energy within the reconnection layer, the temperature of the layer decreases. To maintain pressure equilibrium in the region, the density within the layer correspondingly increases to offset the lower temperature. This compression, in turn, increases the radiation, resulting in a feedback loop of runaway compression and cooling of the layer.

The relationship between plasmoid formation and plasma instabilities has been explored in previous studies \citep[e.g.,][]{2012PhPl...19d2303L,2017ApJ...841...27N}.
The Kelvin-Helmholtz (KH) instability is one mechanism that facilitates plasmoid formation at the shear boundary between the jet funnel and sheath regions. The tearing instability, on the other hand, is commonly associated with magnetic reconnection through an extended thin current sheet, resulting in plasmoid chains. \cite{2022ApJ...924L..32R} examined the development of plasmoid chains in the equatorial plane under the MAD state. Similar plasmoid chain formation is observed in our simulations. In this study, we investigate the properties of these plasmoids in greater detail.

\begin{figure*}
    \centering
    \includegraphics[width=0.8\linewidth]{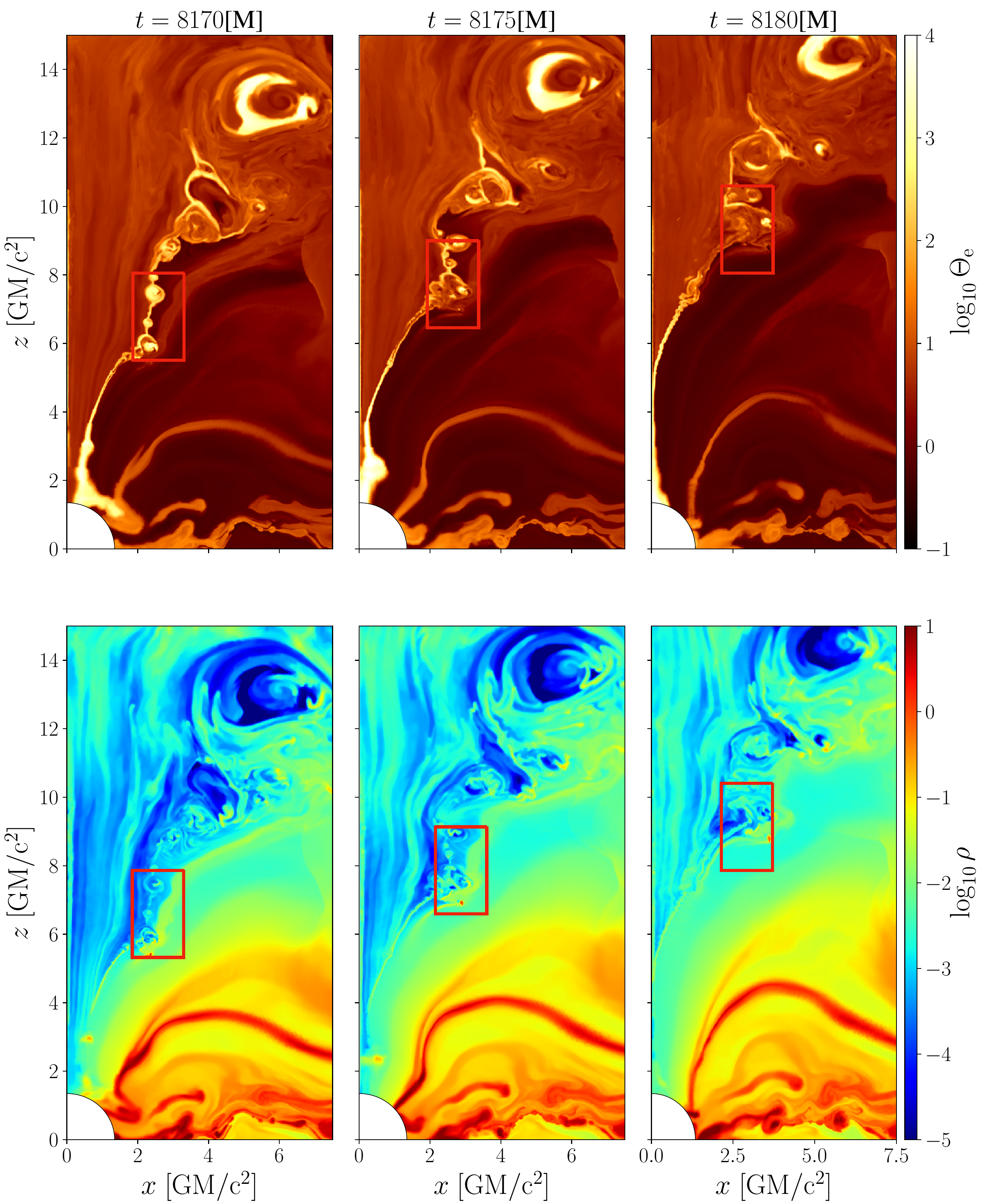}
    \caption{
    Time evolution of a plasmoid chain in high-resolution model {\tt hr} over the time interval $t = 8170\,\rm M$ to $t = 8180\,\rm M$. The upper panel shows electron temperature, and the lower panel presents density. We mark the location of the plasmoid chain with a red box in each panel.
    }
    \label{fig:Fig7}
\end{figure*}

\begin{figure}
    \centering
    \includegraphics[width=\linewidth]{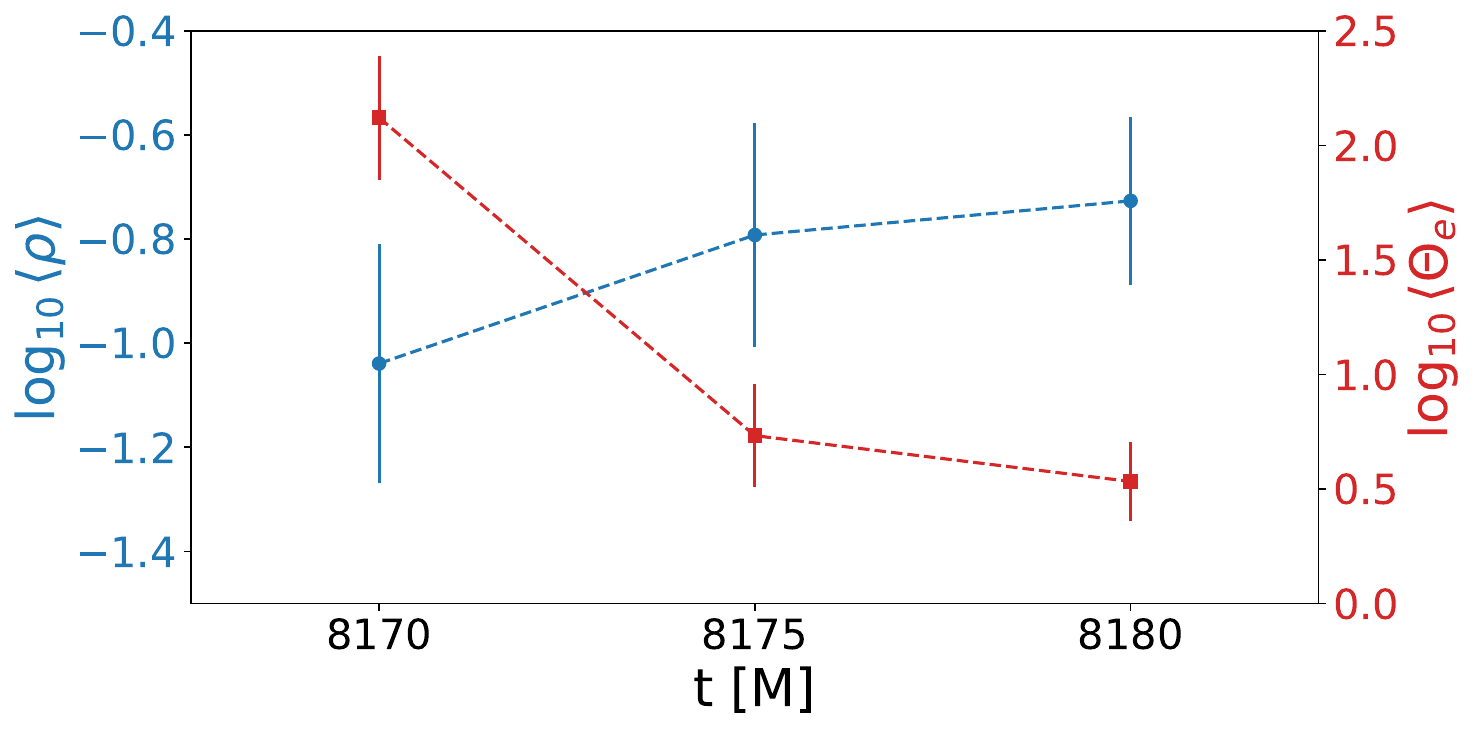}
    \caption{
    Time evolution of integrated density (blue) and electron temperature (red) of tracked plasmoid in the same interval as shown in Fig.~\ref{fig:Fig7}. 
    }
    \label{fig:rho_Te}
\end{figure}

As radiative losses deplete internal energy from the reconnection layer, its temperature decreases. In response, the density increases to maintain pressure equilibrium, as previously discussed, ultimately leading to collapse.
In addition to radiative collapse, the reconnection layer may also be subject to local radiative cooling instabilities \citep{1965ApJ...142..531F}. These instabilities arise from pressure perturbations that become unstable under radiative cooling. The interaction between these thermal and the tearing instabilities can play a significant role in shaping the transient dynamics of the reconnection process \citep{1991SoPh..135..361F}. In our simulations, we observe a collapse induced by radiative cooling.

As shown in upper panels of Fig.~\ref{fig:Fig7}, we focus on the evolution of a single plasmoid chain in high-resolution strong cooling model $\tt{hr}$, which effectively illustrates the process of radiative collapse under radiative cooling. As predicted, the plasmoids in the chain become cooler (dimmer) as they evolve outward, ultimately collapsing under strong compression.
Additionally, we check the density distribution around the plasmoid chain shown in the lower panel in Fig.~\ref{fig:Fig7}. 
The temporal evolution of integrated electron temperature and density in the vicinity of plasmoids is shown in Fig.~\ref{fig:rho_Te}.
It is evident that, as the plasmoid chain evolves, the surrounding electron temperature decreases while density gradually increases, 
explaining why the plasmoids eventually undergo radiative collapse. Similar phenomena are also observed in the lower-resolution simulations, as shown in Fig.~\ref{fig:lr}.

\subsubsection{Statistical properties of plasmoid}
\label{detect}

In order to analyze the statistical properties of plasmoids with radiative cooling, we aim to detect them in the simulations. We apply a plasmoid detection method based on \cite{2020MNRAS.495.1549N}, by tracking blob-like structures in our simulations using the scikit-image package \citep{van2014scikit}. This approach allows us to capture the positions and sizes of the plasmoids within a selected region, which extends from $0$ to $20\,r_{\rm g}$ in $x$-direction and from  $-20$ to $20\,r_{\rm g}$ in $y$-direction of the snapshots with a cadence of  $10\,\rm M$. We applied the same detection criteria to all models to ensure generalizability. We also constrain the regions where the Bernoulli constant ($-hu_t$) is greater than 1.02 for this plasmoid detection. 

\begin{figure}
    \centering
    \includegraphics[width=\linewidth]{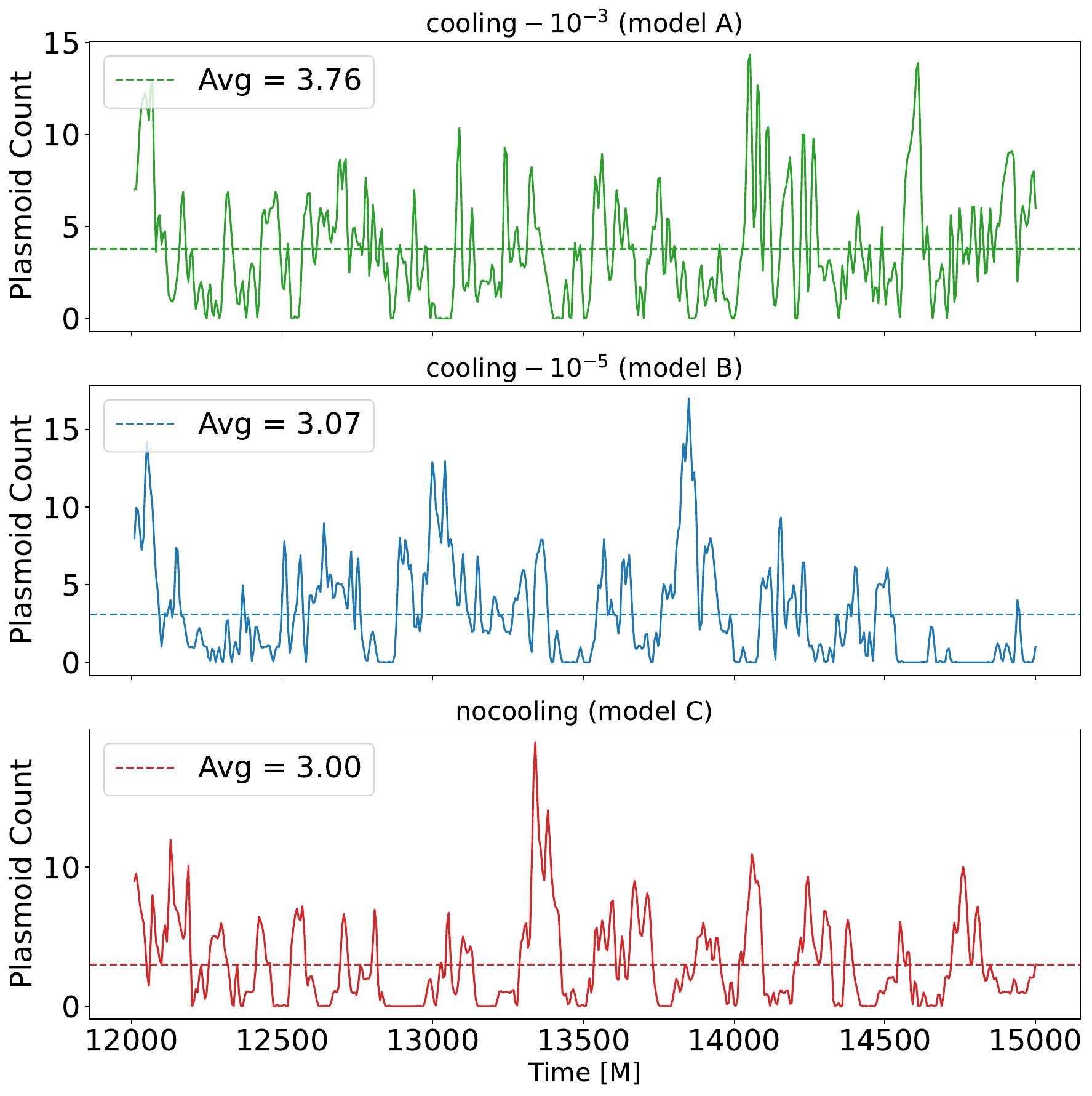}
    \caption{
    Time evolution of number counts of detected plasmoid in different models, (top) cooling with $\dot{M}/\dot{M}_{\rm Edd}=10^{-3}$, (middle) cooling with $\dot{M}/\dot{M}_{\rm Edd}=10^{-5}$, (low) non-cooing.
    }
    \label{fig:Fig10}
\end{figure}

As discussed in Sect.~\ref{collaspe}, radiative cooling leads to the collapse of the plasmoids. This implies that the number of detected plasmoids decreases after the collapse occurs.
In Fig.~\ref{fig:Fig10}, we plot the time evolution of number of plasmoids in different models. 
In the presence of radiative cooling, radiative collapse causes the plasmoid to dissipate and gradually vanish. Compared to cases without radiative cooling, this collapse shortens the lifetime of the plasmoid chain. This results in a more spiky structure in the plasmoid count evolution.

In our ideal GRMHD framework (i.e., physical resistivity $\eta = 0$), all resistive effects arise from numerical dissipation. From our second-order numerical accuracy of the \texttt{BHAC} code \citep{2017ComAC...4....1P}, the effective numerical resistivity scales as $\eta_{\rm num} \sim \Delta x^p$ with $p \approx 2$. As a result, the Lundquist number, defined as $S_L = V_A L / \eta$, remains independent of local temperature variations.Here, $V_A$ and $L$ denote the Alfv\'en speed and the size of the reconnection layer, respectively. 
The normalized reconnection rate $\sim A^{1/2} S_L^{-1/2}$, mainly depends on the density compression ratio $A \equiv \rho_{\text{layer}} / \rho_{\text{in}}$  \citep{2011PhPl...18d2105U}. In our simulations, the radiative cooling leads to compression and density enhancement near the reconnection layer, as illustrated in the red box in Fig.~\ref{fig:Fig7}. 
It increases the compression ratio and leads to a higher magnetic reconnection rate. We present the time evolution of detected plasmoids in Fig.~\ref{fig:Fig10}, which shows more frequent production of plasmoid chains in the cooling case (see the top panel of Fig.~\ref{fig:Fig10}).

\begin{figure}
    \centering
    \includegraphics[width=\linewidth]{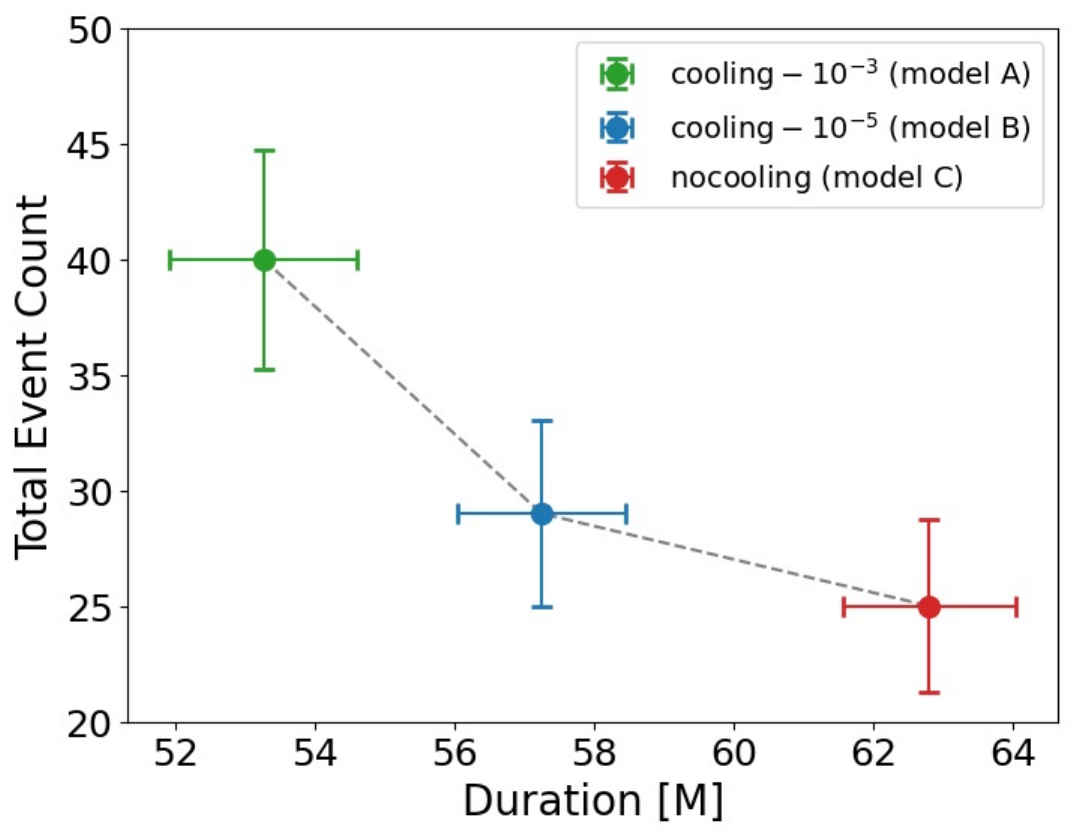}
    \caption{
    This figure shows the average duration and the number of plasmoid events for the three cases. 
    Model A exhibits more frequent magnetic reconnection events, leading to more plasmoid events; however, radiative cooling also causes plasmoids to dissipate earlier, resulting in relatively shorter durations.}
    \label{fig:Duration}
\end{figure}

\begin{figure}
    \centering
    \includegraphics[width=\linewidth]{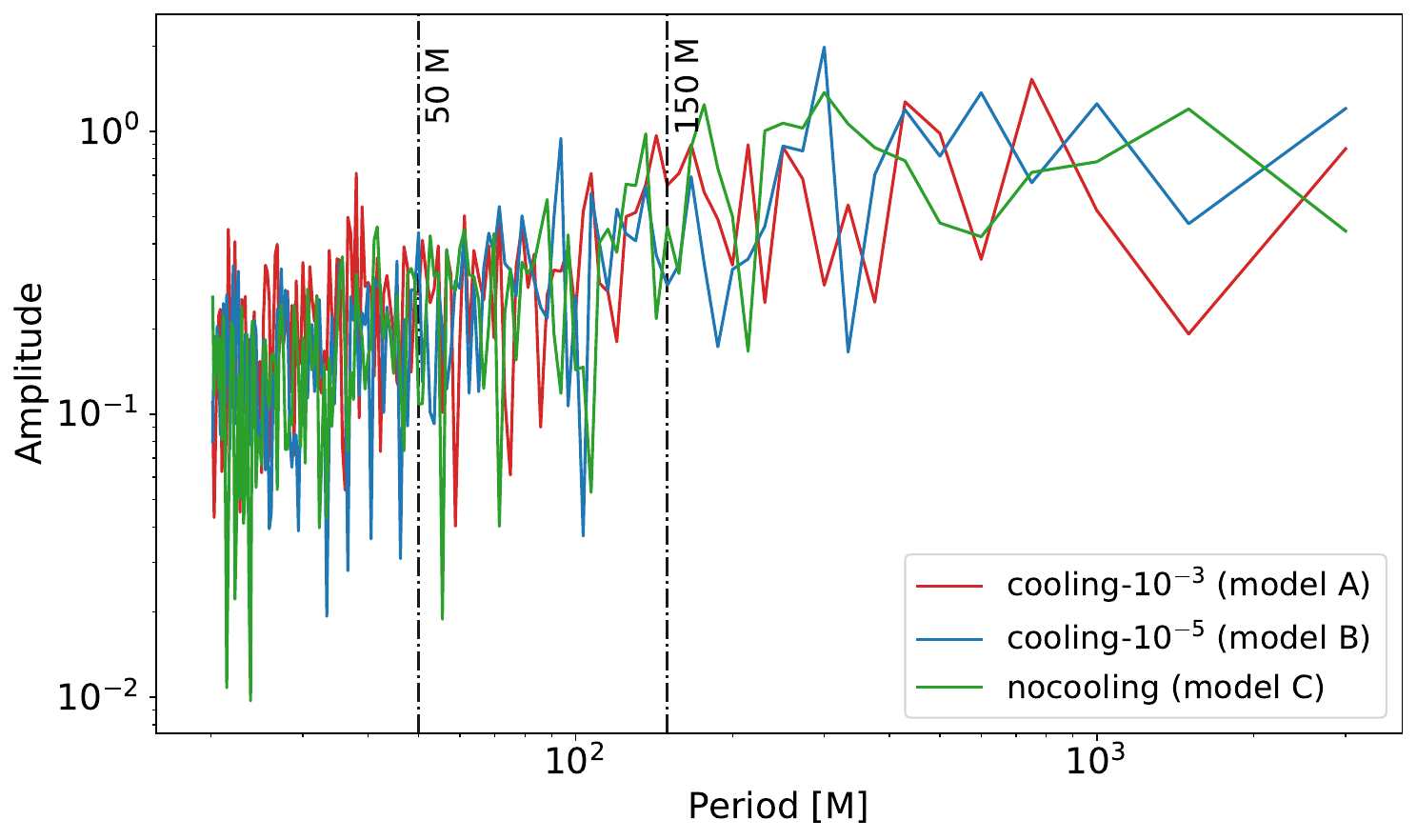}
    \caption{The figure presents the power spectral density (PSD) of the plasmoid counts, obtained via temporal Fourier analysis for different cases. Vertical dashed lines mark the divisions between high (0–-50,$\,\rm M$), mid (50–-150,$\,\rm M$), and low (>150$\,\rm M$) frequency bands.
    }
    \label{fig:plasmoid_fft_with_bands_only}
\end{figure}

To facilitate a more quantitative discussion, we perform a statistical analysis of the transient behaviors of the plasmoid chains, which is shown in Fig.~\ref{fig:Duration}. 
Here, we define the duration based on the criterion of the mean value minus standard deviation. The time points where the count crosses below this threshold are identified as the start and end of individual plasmoid events.
This result aligns with our expectations: radiative collapse shortens the lifetime of the plasmoid chains.
Additionally, the increased density and compressed layer due to radiative cooling further increase the magnetic reconnection rate, yielding more frequent production of plasmoid chains. We note that the average number of plasmoids is not influenced significantly by the cooling effect, which we attribute to the combined effect of these two processes. 
To complement the temporal diagnostics, we conduct a Fourier analysis over the same time interval to investigate variability associated with plasmoid formation, as presented in Fig.~\ref{fig:plasmoid_fft_with_bands_only} and summarize the mean amplitudes in three different frequencies in Table~\ref{tab:freq_amp}. 

In the high-frequency range ($0 - 50,\rm M$), the radiatively cooled case (model A) shows a higher average amplitude, indicating the occurrence of rapid reconnection events and radiatively driven plasmoid collapses.
In contrast, the non-cooling model (model C) shows the highest average amplitude in the low-frequency range (greater than $150\,\rm M$), suggesting a dominant contribution from the longer lifetime of plasmoid chains.
As the strength of radiative cooling increases (from model B to model A), the average amplitude in the low-frequency band gradually decreases, implying that radiative collapse significantly shortens the duration of large-scale plasmoid events (see Table~\ref{tab:freq_amp}).

\begin{table}
\caption{Mean amplitudes of plasmoid fluctuations in different frequency bands.}
\label{tab:freq_amp}
\centering
\begin{tabular}{lccc}
\hline\hline
Model & High Frequency & Mid Frequency & Low Frequency \\
\hline
model C       & 0.15672 & 0.32178 & 0.79337 \\
model B       & 0.17151 & 0.32264 & 0.74936 \\
model A       & 0.21920 & 0.34645 & 0.66034 \\
\hline
\end{tabular}
\end{table}

\begin{figure}
    \centering
    \includegraphics[width=\linewidth]{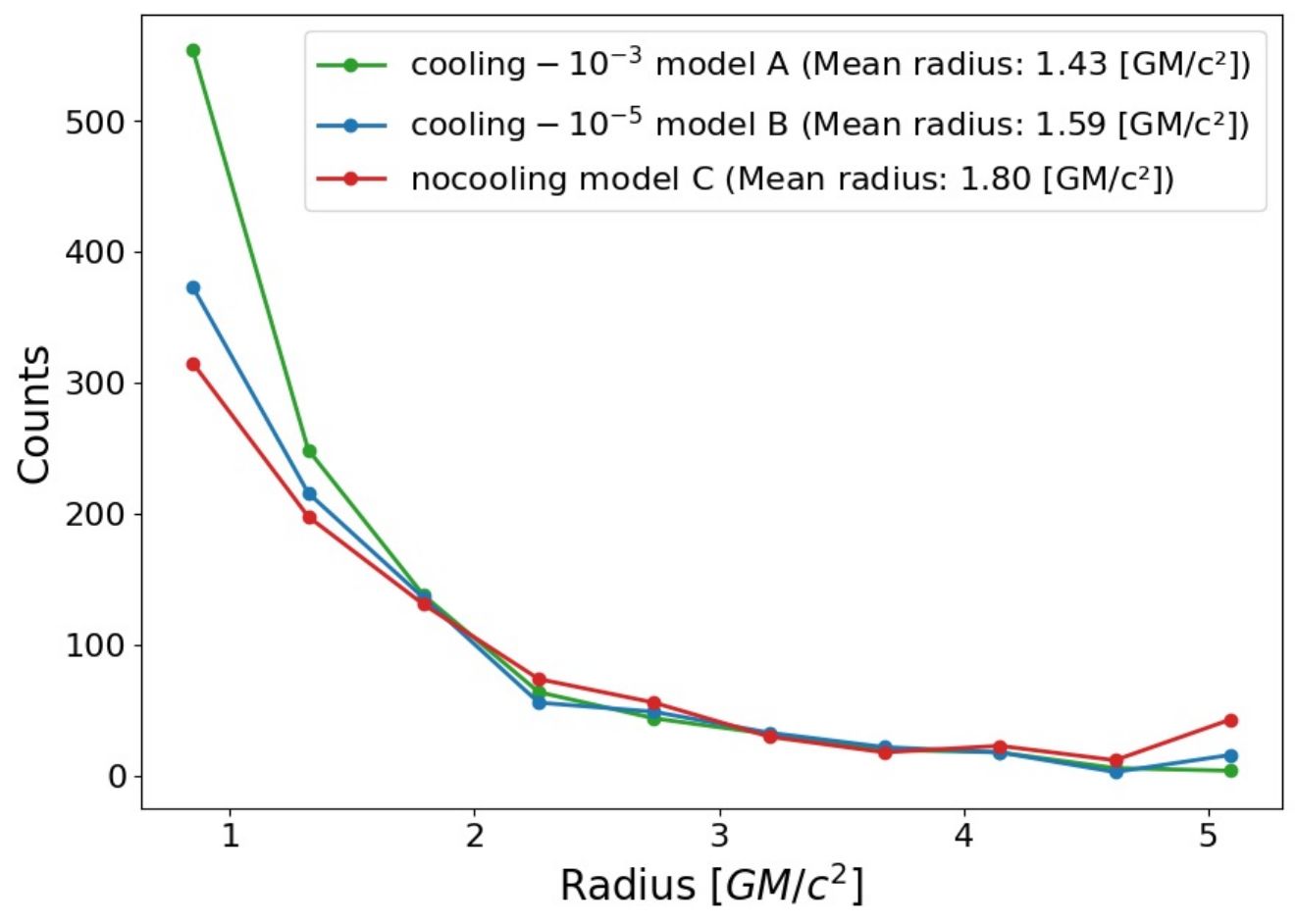}
    \caption{The number counts of the detected plasmoid radii in different models (green: cooling with $\dot{M}/\dot{M}_{\rm Edd}=10^{-3}$, blue: cooling with $\dot{M}/\dot{M}_{\rm Edd}=10^{-5}$, red: non-cooling).
    }
    \label{fig:Fig12}
\end{figure}

To further explore the impact of radiative cooling on the plasmoids, we analyze and compare the radii of plasmoids detected in models A, B, and C, shown in Fig.~\ref{fig:Fig12}. The results reveal that the layer compression caused by radiative cooling also affects plasmoids, resulting in a tendency toward the production of smaller plasmoids. The mean radius of Models A, B, and C is $r=1.43$, $1.59$, and $1.80\,\rm G \rm M/\rm c^2$. It is reflected in the increase of the total number of plasmoids seen in Fig.~\ref{fig:Fig10}.

\subsubsection{Magnetic field accumulation}\label{flux}

Magnetic flux pile-up occurs when the rate of magnetic flux injection, $\tau_{\text{inj}}^{-1} / \tau_A^{-1} \sim V_{\text{in}} / V_{A,1} \equiv M_{A,1}$, exceeds that of flux annihilation in the reconnection layer, $ \tau_R^{-1} / \tau_A^{-1}$ \citep{1986PhFl...29.1520B}. Here, $\tau_{\text{inj}}^{-1}$ and $\tau_R^{-1}$ are the flux injection and reconnection rates, respectively, $\tau_A$ is the Alfvén transit time, $V_{\text{in}}$ is the inflow velocity, and $V_{A,1}$ and $M_{A,1}$ are the Alfvén velocity and Mach number in the inflow, respectively.

Due to the increased reconnection rate, as discussed in Sect .~\ref{detect}, a greater amount of magnetic flux undergoes annihilation along the current sheet, thereby reducing magnetic flux $\Phi_\text{BH}$.
This explains the trend in magnetic flux variation observed in Fig.~\ref{fig:Fig1}(d), where stronger radiative cooling results in less magnetic flux accumulation.
For larger magnetic loops, such as in models D, E, and F, we expect more magnetic field accumulation at the event horizon. As discussed in Sect.~\ref{subsec:accretion}, radiative cooling suppresses the development of the MAD state. For instance, in model D, strong radiative cooling case, the magnetic field fails to reach the critical strength required to choke the accretion flow, and as a result, the MAD state does not emerge.
Similarly, the weak radiative cooling in model E delays its entry into the MAD state compared to model F.
Furthermore, the duration of each periodic MAD phase of the weak cooling case is noticeably shorter, which is reflected in sharper sawtooth-like structures in Fig.~\ref{fig:Fig1}(d). We attribute this to the reduced accumulation of magnetic fields (stronger magnetic flux annihilation) caused by radiative cooling. When the magnetic flux meets the threshold for the MAD state, the reduction in magnetic field strength quickly returns it to the SANE state soon.

\subsubsection{Energy Extraction} \label{EE}

Plasmoid chain created by magnetic reconnection in BH ergosphere is one of the possible mechanisms of energy extraction from a rotating BH via the Penrose process (PP) \citep[e.g.,][]{2015PhRvL.114k5003A}.
In previous studies of reconnection-driven PP, the discussion has primarily focused on current sheets forming near the equatorial plane of the BH \citep{2008ApJ...682.1124K,2015PhRvL.114k5003A}. A recent study of \cite{2025ApJ...982L..31C} has shown that PP can also occur outside the equatorial plane, as strong magnetic reconnection regions are present on the transition region between the accretion disk and the jet \citep{2025A&A...696A..36D}. To precisely characterize PP, a local model of magnetic reconnection is required, providing a prescription for the plasmoid outflow velocity field $\tilde{v}_{\text{out}}$ and their associated energetics. However, the quantities of greatest importance, outflow velocity and the reconnection rate, come from MHD approximation \citep{2025PhRvD.111b3003S,2025ApJ...982L..31C}. In our study, we do not pursue further theoretical analysis. Instead, we focus on how radiative cooling impact the energy extraction process via magnetic reconnection although it happens numerically due to numerical resistivity in ideal GRMHD simulations.

We then follow the definition provided in \cite{2025PhRvD.111b3003S}, the energy consevation $\nabla_\mu T^\mu_t=0$, leading to the following form:
\begin{equation}
\partial_t e + \frac{1}{h_r h_\theta h_\phi} \partial_i \left(h_r h_\theta h_\phi S^i\right) = 0.
\end{equation}
where $e=-\alpha T^t_t$ denotes the so-called energy-at-infinity density. Here, \(h_i = \sqrt{g_{ii}}\) denotes the scale factor in the context of the \(3+1\) formalism, where \(g_{ii}\) represents the metric components in Boyer–-Lindquist coordinates.

\begin{figure}
    \centering
    \includegraphics[width=0.4\textwidth]{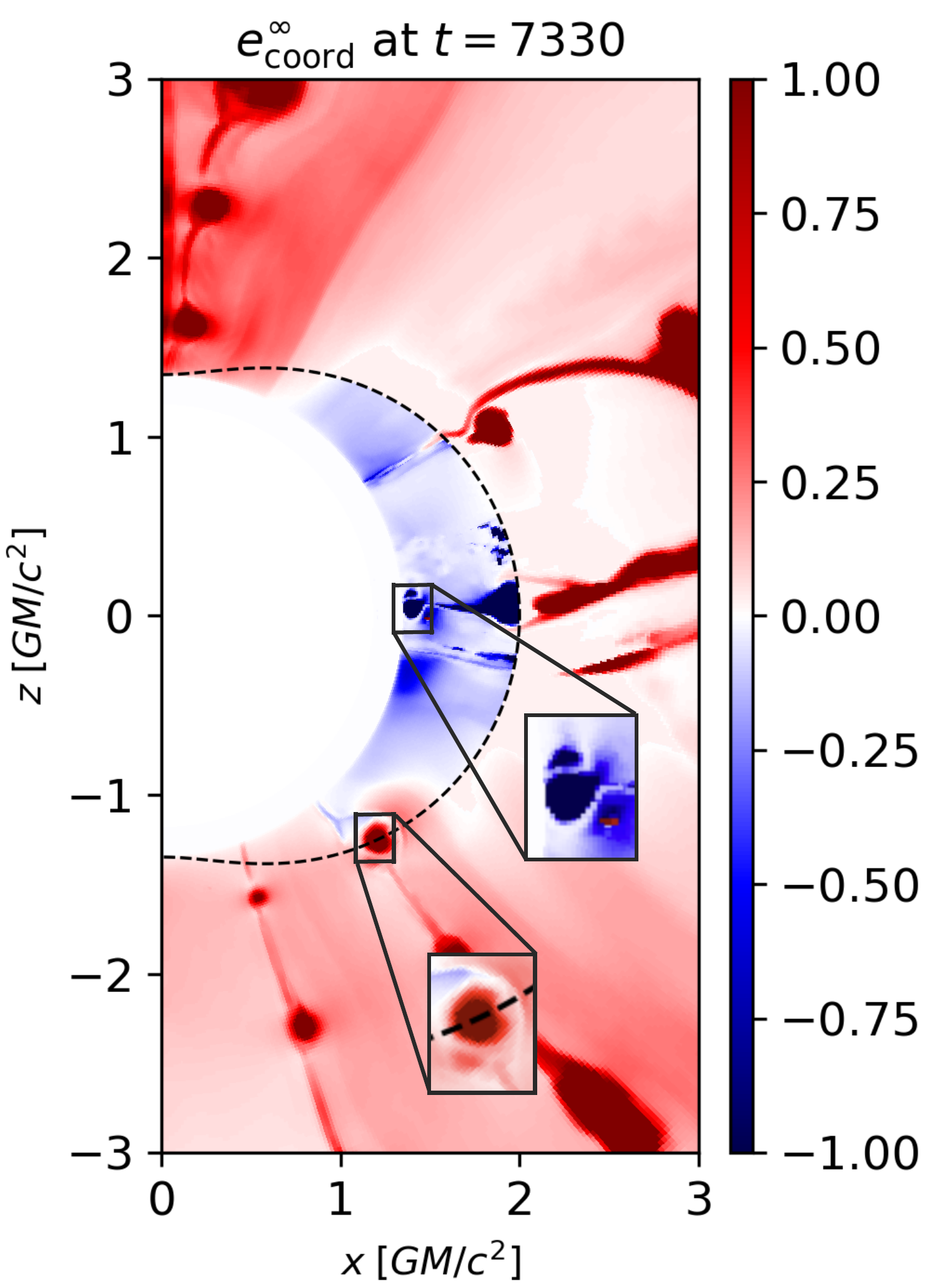}
    \caption{
    The distribution of the normalized energy-at-infinity density $e^\infty$ for model A within $3~r_g$ at $t=7330\,\rm M$. The dashed line indicates the position of ergosphere.
    The zoomed-in region shows two plasmoids: one associated with negative energy falling into the BH, and the other with positive energy escaping outward. 
    }
    \label{fig:pp}
\end{figure}

Fig.~\ref{fig:pp} shows the distribution of energy-at-infinity density near the BH from our simulation of Model A at \(t = 7330\,\rm M\). As shown in the zoomed-in region, a negative energy density structure is embedded within a plasmoid. This feature arises from the tearing instability of the current sheet and subsequent magnetic reconnection near the plasmoid, leading to the formation of both negative and positive energy regions \citep{2025ApJ...982L..31C}. One plasmoid carries negative energy and falls into the BH, while another carries positive energy and escapes outward. This bidirectional energy flow reflects a Penrose-like mechanism driven by reconnection. We note that only the escaping positive energy flux contributes to extracting energy from the BH, whereas negative energy falling inward contributes to the total energy budget without aiding extraction.

\begin{figure}
    \centering
    \includegraphics[width=0.35\textwidth]{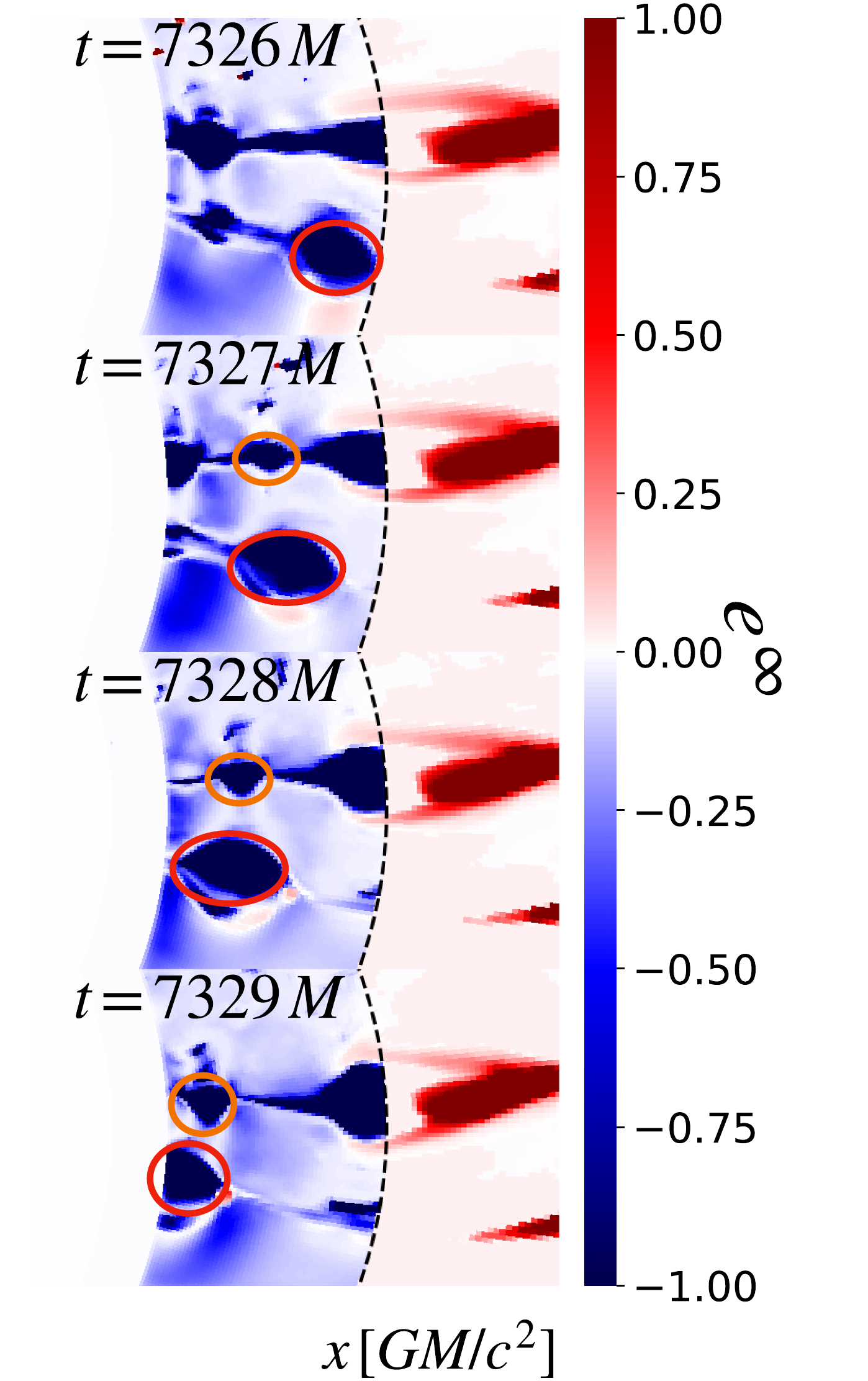}
    \caption{
    Time evolution of the normalized energy-at-infinity density near the horizon around the equatorial plane for model A. The plasmoid with negative energy-at-infinity falls into the BH near the equatorial plane. The regions outlined in red and orange highlight two plasmoid chains with inflowing negative energy-at-infinity. The dashed lines indicate the position of the ergosphere.
    }
    \label{fig:pp-evo}
\end{figure}

To present directly that the plasmoid carrying negative energy-at-infinity  falls into the BH, we track the evolution of a selected plasmoid near the equatorial plane, as shown in Fig.~\ref{fig:pp-evo}. The time evolution clearly shows the inward motion of the plasmoid with negative energy-at-infinity as it plunges into the BH.

\begin{figure}
    \centering
    \includegraphics[width=0.47\textwidth]{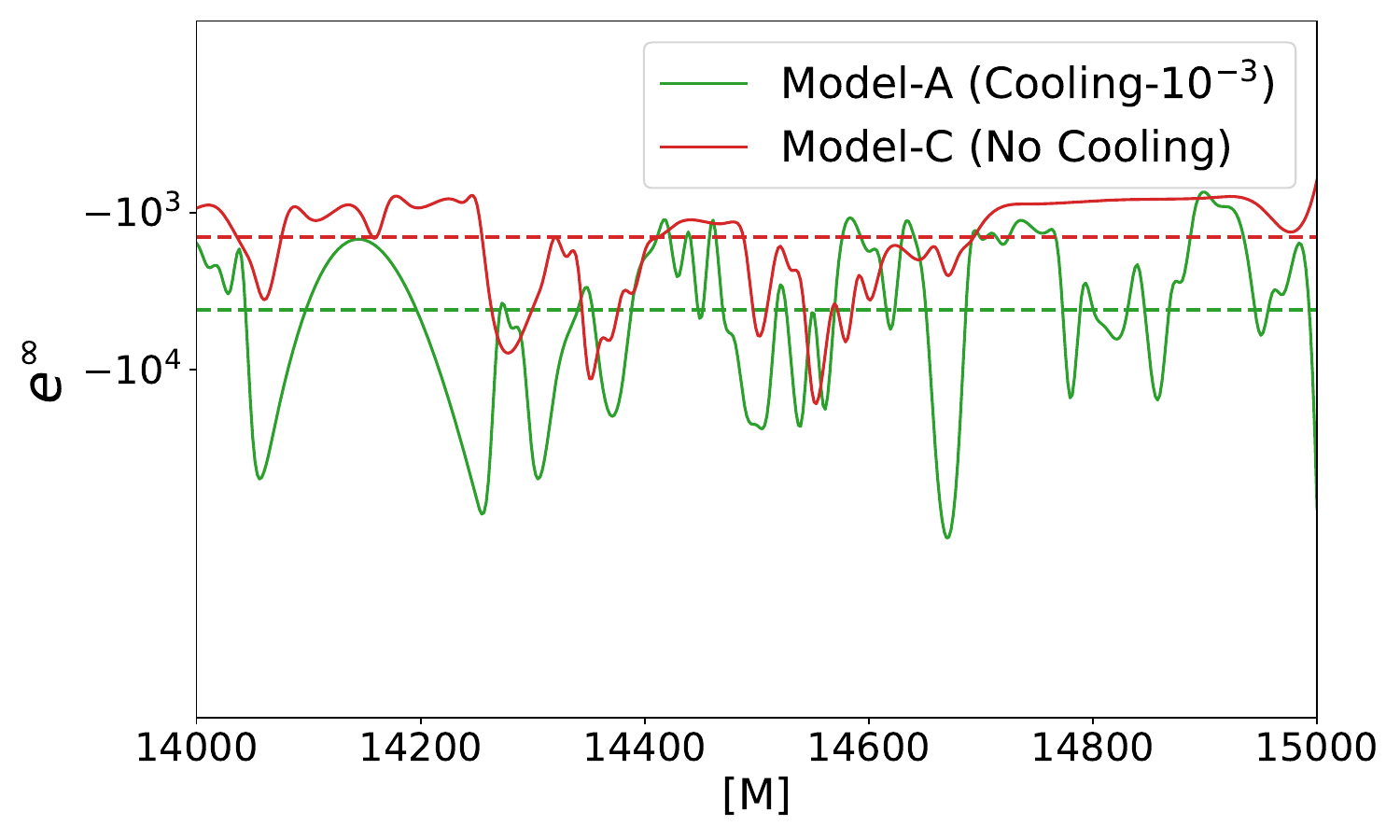}
    \caption{
    The figure presents the differences in the total negative energy among the different models. The horizontal dashed lines denote the average value of each model.
    }
    \label{fig:negen}
\end{figure}

As discussed in Sect.~\ref{detect}, radiative cooling affects the local reconnection rate and the efficiency of plasmoid formation. As shown in Fig.~\ref{fig:negen}, we adopt previously introduced \(-hu_t\) as the diagnostic criterion to identify negative-energy regions within plasmoids, and compute the total negative energy by summing over these regions for each model. 
In model A (strong radiative cooling), two prominent peaks are observed around $t = 14050\,\rm M$ and $t = 14700\,\rm M$. 
Similarly, in Fig.~\ref{fig:Fig10}, peaks corresponding to plasmoid formation events are observed near these times, further supporting the temporal correlation. 
For Model C (no cooling), although no prominent peaks are observed, there still exists noticeable variations corresponding to plasmoid formation events between $t = 14200\,\rm M$ and $t = 14600\,\rm M$, compared to the surrounding periods. Model A exhibits a higher average negative energy-at-infinity density compared to the non-cooling model C. 
Thus, to some extent, radiative cooling may lead to more efficient energy extraction. Although a rigorous calculation of energy extraction requires a more detailed discussion, the trend-based analysis of negative energy differences across different models presented here provide supporting evidence for our discussion on the effects of radiative cooling.

\section{Summary and discussion}\label{conclusion}

We performed a series of 2D and 3D two-temperature GRMHD simulations initialized with a multi-loop configuration to systematically examine the effect of radiative cooling on the accretion flows. We focus on the cooling-induced collapse of plasmoids, analyzing their statistical properties under varying cooling strength. We further investigate how these radiative-cooling driven collapse events alter the small-scale structure of the current sheet and assess their potential impact on energy extraction mechanisms. Below, we summarize our conclusions in pointwise:

\begin{enumerate}
    \item Radiative cooling significantly modifies the accretion dynamics near the BH, reducing the magnetic flux accumulation at the event horizon and influencing the transition between the SANE and MAD states.
    \item Radiative cooling globally reduces the dimensionless electron temperature and increases the disk density through enhanced compression and alterations in the magnetic field structure, leading to observable modifications in the disk morphology.
    \item Cooling induces layer compression within the reconnection regions, resulting in smaller plasmoid structures and promoting faster radiative collapse, thereby shortening the lifetime of individual plasmoid events.
    \item Radiative cooling results in a higher frequency of plasmoid formation. However, the average ``lifetime'' of individual plasmoids decreases due to cooling-induced collapse. The radiative cooling case tends to have more small-scale reconnection events and radiative plasmoid collapses. The interplay between enhanced reconnection and rapid collapse maintains a similar average number of plasmoids.
    \item A comparison of negative energy-at-infinity density variations under different radiative cooling conditions reveals the presence of peaks associated with plasmoid formation events. Radiative cooling may modulate the efficiency and characteristics of energy extraction to some extent.
\end{enumerate}

Recent observations of rapid variability in X-ray and gamma-ray flares from AGNs, with timescales ranging from several minutes to a few hours, impose stringent constraints on particle acceleration timescales and the size of the emission region \citep{2007A&A...470..475A,2007ApJ...663..125A}. Observational evidence suggests that these fast-variable flares originate from a compact region with a size on the order of a few Schwarzschild radii.
Our results suggest that the collapse induced by radiative cooling confines plasmoid chains to smaller spatial regions and shortens their temporal duration. Therefore, our radiative cooling simulations may, to some extent, correspond to the observed variability in AGNs. The trends revealed by our results also offer a potential explanation for low accretion rate systems, such as Sgr~A$^*$.

Owing to limited computational resources, the presence of plasmoids cannot be resolved in our current 3D simulations, preventing us from modeling and tracking their temporal evolution. In future work, we plan to explore the impact of radiative cooling on the accretion process using higher-resolution, multi-loop 3D GRMHD simulations, with the goal of detecting and characterizing plasmoids in fully three-dimensional settings. Moreover, the cooling model implemented in our simulations corresponds to an accretion rate substantially higher than that inferred for Sgr~A$^*$. 
At such accretion rates, of order $\dot{M} \sim 10^{-3}\,\dot{M}_{\rm Edd}$, additional radiative processes---most notably inverse Compton scattering---as well as radiation back-reaction on the plasma dynamics may become dynamically important. 
Since these effects are not yet included in a fully self-consistent manner in our current cooling prescription, the quantitative details of the accretion-flow structure, plasmoid evolution, and energy-extraction efficiency may be affected. 
Consequently, our results should be regarded as indicative of qualitative trends rather than precise quantitative predictions at these accretion rates.
In future studies, we plan to construct a more self-consistent cooling framework tailored to the physical conditions of Sgr~A$^*$, with the goal of better understanding its observed flaring behavior.

\begin{acknowledgements}
This research was supported by the National Key Research and Development Program of China (grant no. 2023YFE0101200), the National Natural Science Foundation of China (grant nos. 12273022, 12511540053), and the Shanghai Municipality Orientation Program of Basic Research for International Scientists (grant no. 22JC1410600). C.M.F. is supported by the DFG research grant “Jet physics on horizon scales and beyond” (Grant No. 443220636) within the DFG research unit “Relativistic Jets in Active Galaxies” (FOR 5195). The simulations were performed on the TDLI-Astro cluster at Shanghai Jiao Tong University.
\end{acknowledgements}

\bibliography{aa}

@ARTICLE{2022MNRAS.511.3795N,
       author = {{Narayan}, Ramesh and {Chael}, Andrew and {Chatterjee}, Koushik and {Ricarte}, Angelo and {Curd}, Brandon},
        title = "{Jets in magnetically arrested hot accretion flows: geometry, power, and black hole spin-down}",
      journal = {\mnras},
         year = 2022,
        month = apr,
       volume = {511},
       number = {3},
        pages = {3795-3813},
          doi = {10.1093/mnras/stac285},
archivePrefix = {arXiv},
       eprint = {2108.12380},
 primaryClass = {astro-ph.HE},
       adsurl = {https://ui.adsabs.harvard.edu/abs/2022MNRAS.511.3795N}
}

@ARTICLE{2024ApJ...974..116C,
       author = {{Chavez}, Erandi and {Issaoun}, Sara and {Johnson}, Michael D. and {Tiede}, Paul and {Fromm}, Christian and {Mizuno}, Yosuke},
        title = "{Prospects of Detecting a Jet in Sagittarius A* with Very-long-baseline Interferometry}",
      journal = {\apj},
         year = 2024,
        month = oct,
       volume = {974},
       number = {1},
          eid = {116},
        pages = {116},
          doi = {10.3847/1538-4357/ad6b1f},
archivePrefix = {arXiv},
       eprint = {2405.06029},
 primaryClass = {astro-ph.HE},
       adsurl = {https://ui.adsabs.harvard.edu/abs/2024ApJ...974..116C}
}

@ARTICLE{2017ComAC...4....1P,
       author = {{Porth}, Oliver and {Olivares}, Hector and {Mizuno}, Yosuke and {Younsi}, Ziri and {Rezzolla}, Luciano and {Moscibrodzka}, Monika and {Falcke}, Heino and {Kramer}, Michael},
        title = "{The black hole accretion code}",
      journal = {Computational Astrophysics and Cosmology},
         year = 2017,
        month = may,
       volume = {4},
       number = {1},
          eid = {1},
        pages = {1},
          doi = {10.1186/s40668-017-0020-2},
archivePrefix = {arXiv},
       eprint = {1611.09720},
 primaryClass = {gr-qc},
       adsurl = {https://ui.adsabs.harvard.edu/abs/2017ComAC...4....1P}
}

@ARTICLE{2019A&A...629A..61O,
       author = {{Olivares}, Hector and {Porth}, Oliver and {Davelaar}, Jordy and {Most}, Elias R. and {Fromm}, Christian M. and {Mizuno}, Yosuke and {Younsi}, Ziri and {Rezzolla}, Luciano},
        title = "{Constrained transport and adaptive mesh refinement in the Black Hole Accretion Code}",
      journal = {\aap},
         year = 2019,
        month = sep,
       volume = {629},
          eid = {A61},
        pages = {A61},
          doi = {10.1051/0004-6361/201935559},
archivePrefix = {arXiv},
       eprint = {1906.10795},
 primaryClass = {astro-ph.HE},
       adsurl = {https://ui.adsabs.harvard.edu/abs/2019A&A...629A..61O}
}

@ARTICLE{2015GReGr..47....3G,
       author = {{Giulini}, Domenico},
        title = "{Luciano Rezzolla and Olindo Zanotti: Relativistic hydrodynamics. Oxford University Press, Oxford, 2013, 752 pp, GBP 55.00, ISBN: 978-0-19-852890-6}",
      journal = {General Relativity and Gravitation},
         year = 2015,
        month = feb,
       volume = {47},
          eid = {3},
        pages = {3},
          doi = {10.1007/s10714-014-1839-3},
       adsurl = {https://ui.adsabs.harvard.edu/abs/2015GReGr..47....3G}
}

@ARTICLE{2017MNRAS.466..705S,
       author = {{S{\k{a}}dowski}, Aleksander and {Wielgus}, Maciek and {Narayan}, Ramesh and {Abarca}, David and {McKinney}, Jonathan C. and {Chael}, Andrew},
        title = "{Radiative, two-temperature simulations of low-luminosity black hole accretion flows in general relativity}",
      journal = {\mnras},
         year = 2017,
        month = apr,
       volume = {466},
       number = {1},
        pages = {705-725},
          doi = {10.1093/mnras/stw3116},
archivePrefix = {arXiv},
       eprint = {1605.03184},
 primaryClass = {astro-ph.HE},
       adsurl = {https://ui.adsabs.harvard.edu/abs/2017MNRAS.466..705S}
}

@ARTICLE{2015MNRAS.454.1848R,
       author = {{Ressler}, S.~M. and {Tchekhovskoy}, A. and {Quataert}, E. and {Chandra}, M. and {Gammie}, C.~F.},
        title = "{Electron thermodynamics in GRMHD simulations of low-luminosity black hole accretion}",
      journal = {\mnras},
         year = 2015,
        month = dec,
       volume = {454},
       number = {2},
        pages = {1848-1870},
          doi = {10.1093/mnras/stv2084},
archivePrefix = {arXiv},
       eprint = {1509.04717},
 primaryClass = {astro-ph.HE},
       adsurl = {https://ui.adsabs.harvard.edu/abs/2015MNRAS.454.1848R}
}

@ARTICLE{1976ApJ...207..962F,
       author = {{Fishbone}, L.~G. and {Moncrief}, V.},
        title = "{Relativistic fluid disks in orbit around Kerr black holes.}",
      journal = {\apj},
         year = 1976,
        month = aug,
       volume = {207},
        pages = {962-976},
          doi = {10.1086/154565},
       adsurl = {https://ui.adsabs.harvard.edu/abs/1976ApJ...207..962F}
}

@ARTICLE{2021MNRAS.508.1241C,
       author = {{Chashkina}, Anna and {Bromberg}, Omer and {Levinson}, Amir},
        title = "{GRMHD simulations of BH activation by small scale magnetic loops: formation of striped jets and active coronae}",
      journal = {\mnras},
         year = 2021,
        month = nov,
       volume = {508},
       number = {1},
        pages = {1241-1252},
          doi = {10.1093/mnras/stab2513},
archivePrefix = {arXiv},
       eprint = {2106.15738},
 primaryClass = {astro-ph.HE},
       adsurl = {https://ui.adsabs.harvard.edu/abs/2021MNRAS.508.1241C}
}

@ARTICLE{2022ApJ...924L..32R,
       author = {{Ripperda}, B. and {Liska}, M. and {Chatterjee}, K. and {Musoke}, G. and {Philippov}, A.~A. and {Markoff}, S.~B. and {Tchekhovskoy}, A. and {Younsi}, Z.},
        title = "{Black Hole Flares: Ejection of Accreted Magnetic Flux through 3D Plasmoid-mediated Reconnection}",
      journal = {\apjl},
         year = 2022,
        month = jan,
       volume = {924},
       number = {2},
          eid = {L32},
        pages = {L32},
          doi = {10.3847/2041-8213/ac46a1},
archivePrefix = {arXiv},
       eprint = {2109.15115},
 primaryClass = {astro-ph.HE},
       adsurl = {https://ui.adsabs.harvard.edu/abs/2022ApJ...924L..32R}
}

@ARTICLE{1996ApJ...465..312E,
       author = {{Esin}, Ann A. and {Narayan}, Ramesh and {Ostriker}, Eve and {Yi}, Insu},
        title = "{Hot One-Temperature Accretion Flows around Black Holes}",
      journal = {\apj},
         year = 1996,
        month = jul,
       volume = {465},
        pages = {312},
          doi = {10.1086/177421},
archivePrefix = {arXiv},
       eprint = {astro-ph/9601074},
 primaryClass = {astro-ph},
       adsurl = {https://ui.adsabs.harvard.edu/abs/1996ApJ...465..312E}
}

@ARTICLE{1995ApJ...452..710N,
       author = {{Narayan}, Ramesh and {Yi}, Insu},
        title = "{Advection-dominated Accretion: Underfed Black Holes and Neutron Stars}",
      journal = {\apj},
         year = 1995,
        month = oct,
       volume = {452},
        pages = {710},
          doi = {10.1086/176343},
archivePrefix = {arXiv},
       eprint = {astro-ph/9411059},
 primaryClass = {astro-ph},
       adsurl = {https://ui.adsabs.harvard.edu/abs/1995ApJ...452..710N}
}

@ARTICLE{2019ApJS..243...26P,
       author = {{Porth}, Oliver and {Chatterjee}, Koushik and {Narayan}, Ramesh and {Gammie}, Charles F. and {Mizuno}, Yosuke and {Anninos}, Peter and {Baker}, John G. and {Bugli}, Matteo and {Chan}, Chi-kwan and {Davelaar}, Jordy and {Del Zanna}, Luca and {Etienne}, Zachariah B. and {Fragile}, P. Chris and {Kelly}, Bernard J. and {Liska}, Matthew and {Markoff}, Sera and {McKinney}, Jonathan C. and {Mishra}, Bhupendra and {Noble}, Scott C. and {Olivares}, H{\'e}ctor and {Prather}, Ben and {Rezzolla}, Luciano and {Ryan}, Benjamin R. and {Stone}, James M. and {Tomei}, Niccol{\`o} and {White}, Christopher J. and {Younsi}, Ziri and {Akiyama}, Kazunori and {Alberdi}, Antxon and {Alef}, Walter and {Asada}, Keiichi and {Azulay}, Rebecca and {Baczko}, Anne-Kathrin and {Ball}, David and {Balokovi{\'c}}, Mislav and {Barrett}, John and {Bintley}, Dan and {Blackburn}, Lindy and {Boland}, Wilfred and {Bouman}, Katherine L. and {Bower}, Geoffrey C. and {Bremer}, Michael and {Brinkerink}, Christiaan D. and {Brissenden}, Roger and {Britzen}, Silke and {Broderick}, Avery E. and {Broguiere}, Dominique and {Bronzwaer}, Thomas and {Byun}, Do-Young and {Carlstrom}, John E. and {Chael}, Andrew and {Chatterjee}, Shami and {Chen}, Ming-Tang and {Chen}, Yongjun and {Cho}, Ilje and {Christian}, Pierre and {Conway}, John E. and {Cordes}, James M. and {Geoffrey} and {Crew}, B. and {Cui}, Yuzhu and {De Laurentis}, Mariafelicia and {Deane}, Roger and {Dempsey}, Jessica and {Desvignes}, Gregory and {Doeleman}, Sheperd S. and {Eatough}, Ralph P. and {Falcke}, Heino and {Fish}, Vincent L. and {Fomalont}, Ed and {Fraga-Encinas}, Raquel and {Freeman}, Bill and {Friberg}, Per and {Fromm}, Christian M. and {G{\'o}mez}, Jos{\'e} L. and {Galison}, Peter and {Garc{\'\i}a}, Roberto and {Gentaz}, Olivier and {Georgiev}, Boris and {Goddi}, Ciriaco and {Gold}, Roman and {Gu}, Minfeng and {Gurwell}, Mark and {Hada}, Kazuhiro and {Hecht}, Michael H. and {Hesper}, Ronald and {Ho}, Luis C. and {Ho}, Paul and {Honma}, Mareki and {Huang}, Chih-Wei L. and {Huang}, Lei and {Hughes}, David H. and {Ikeda}, Shiro and {Inoue}, Makoto and {Issaoun}, Sara and {James}, David J. and {Jannuzi}, Buell T. and {Janssen}, Michael and {Jeter}, Britton and {Jiang}, Wu and {Johnson}, Michael D. and {Jorstad}, Svetlana and {Jung}, Taehyun and {Karami}, Mansour and {Karuppusamy}, Ramesh and {Kawashima}, Tomohisa and {Keating}, Garrett K. and {Kettenis}, Mark and {Kim}, Jae-Young and {Kim}, Junhan and {Kim}, Jongsoo and {Kino}, Motoki and {Koay}, Jun Yi and {Patrick} and {Koch}, M. and {Koyama}, Shoko and {Kramer}, Michael and {Kramer}, Carsten and {Krichbaum}, Thomas P. and {Kuo}, Cheng-Yu and {Lauer}, Tod R. and {Lee}, Sang-Sung and {Li}, Yan-Rong and {Li}, Zhiyuan and {Lindqvist}, Michael and {Liu}, Kuo and {Liuzzo}, Elisabetta and {Lo}, Wen-Ping and {Lobanov}, Andrei P. and {Loinard}, Laurent and {Lonsdale}, Colin and {Lu}, Ru-Sen and {MacDonald}, Nicholas R. and {Mao}, Jirong and {Marrone}, Daniel P. and {Marscher}, Alan P. and {Mart{\'\i}-Vidal}, Iv{\'a}n and {Matsushita}, Satoki and {Matthews}, Lynn D. and {Medeiros}, Lia and {Menten}, Karl M. and {Mizuno}, Izumi and {Moran}, James M. and {Moriyama}, Kotaro and {Moscibrodzka}, Monika and {M{\"u}ller}, Cornelia and {Nagai}, Hiroshi and {Nagar}, Neil M. and {Nakamura}, Masanori and {Narayanan}, Gopal and {Natarajan}, Iniyan and {Neri}, Roberto and {Ni}, Chunchong and {Noutsos}, Aristeidis and {Okino}, Hiroki and {Oyama}, Tomoaki and {{\"O}zel}, Feryal and {Palumbo}, Daniel C.~M. and {Patel}, Nimesh and {Pen}, Ue-Li and {Pesce}, Dominic W. and {Pi{\'e}tu}, Vincent and {Plambeck}, Richard and {PopStefanija}, Aleksandar and {Preciado-L{\'o}pez}, Jorge A. and {Psaltis}, Dimitrios and {Pu}, Hung-Yi and {Ramakrishnan}, Venkatessh and {Rao}, Ramprasad and {Rawlings}, Mark G. and {Raymond}, Alexander W. and {Ripperda}, Bart and {Roelofs}, Freek and {Rogers}, Alan and {Ros}, Eduardo and {Rose}, Mel and {Roshanineshat}, Arash and {Rottmann}, Helge and {Roy}, Alan L. and {Ruszczyk}, Chet and {Rygl}, Kazi L.~J. and {S{\'a}nchez}, Salvador and {S{\'a}nchez-Arguelles}, David and {Sasada}, Mahito and {Savolainen}, Tuomas and {Schloerb}, F. Peter and {Schuster}, Karl-Friedrich and {Shao}, Lijing and {Shen}, Zhiqiang and {Small}, Des and {Sohn}, Bong Won and {SooHoo}, Jason and {Tazaki}, Fumie and {Tiede}, Paul and {Tilanus}, Remo P.~J. and {Titus}, Michael and {Toma}, Kenji and {Torne}, Pablo and {Trent}, Tyler and {Trippe}, Sascha and {Tsuda}, Shuichiro and {van Bemmel}, Ilse and {van Langevelde}, Huib Jan and {van Rossum}, Daniel R. and {Wagner}, Jan and {Wardle}, John and {Weintroub}, Jonathan and {Wex}, Norbert and {Wharton}, Robert and {Wielgus}, Maciek and {Wong}, George N. and {Wu}, Qingwen and {Young}, Ken and {Young}, Andr{\'e} and {Yuan}, Feng and {Yuan}, Ye-Fei and {Zensus}, J. Anton and {Zhao}, Guangyao and {Zhao}, Shan-Shan and {Zhu}, Ziyan and {Event Horizon Telescope Collaboration}},
        title = "{The Event Horizon General Relativistic Magnetohydrodynamic Code Comparison Project}",
      journal = {\apjs},
         year = 2019,
        month = aug,
       volume = {243},
       number = {2},
          eid = {26},
        pages = {26},
          doi = {10.3847/1538-4365/ab29fd},
archivePrefix = {arXiv},
       eprint = {1904.04923},
 primaryClass = {astro-ph.HE},
       adsurl = {https://ui.adsabs.harvard.edu/abs/2019ApJS..243...26P}
}

@ARTICLE{2011MNRAS.418L..79T,
       author = {{Tchekhovskoy}, Alexander and {Narayan}, Ramesh and {McKinney}, Jonathan C.},
        title = "{Efficient generation of jets from magnetically arrested accretion on a rapidly spinning black hole}",
      journal = {\mnras},
         year = 2011,
        month = nov,
       volume = {418},
       number = {1},
        pages = {L79-L83},
          doi = {10.1111/j.1745-3933.2011.01147.x},
archivePrefix = {arXiv},
       eprint = {1108.0412},
 primaryClass = {astro-ph.HE},
       adsurl = {https://ui.adsabs.harvard.edu/abs/2011MNRAS.418L..79T}
}

@ARTICLE{2023MNRAS.522.2307J,
       author = {{Jiang}, Hong-Xuan and {Mizuno}, Yosuke and {Fromm}, Christian M. and {Nathanail}, Antonios},
        title = "{Two-temperature GRMHD simulations of black hole accretion flows with multiple magnetic loops}",
      journal = {\mnras},
         year = 2023,
        month = jun,
       volume = {522},
       number = {2},
        pages = {2307-2324},
          doi = {10.1093/mnras/stad1106},
archivePrefix = {arXiv},
       eprint = {2304.06230},
 primaryClass = {astro-ph.HE},
       adsurl = {https://ui.adsabs.harvard.edu/abs/2023MNRAS.522.2307J}
}

@ARTICLE{2021MNRAS.506..741M,
       author = {{Mizuno}, Yosuke and {Fromm}, Christian M. and {Younsi}, Ziri and {Porth}, Oliver and {Olivares}, Hector and {Rezzolla}, Luciano},
        title = "{Comparison of the ion-to-electron temperature ratio prescription: GRMHD simulations with electron thermodynamics}",
      journal = {\mnras},
         year = 2021,
        month = sep,
       volume = {506},
       number = {1},
        pages = {741-758},
          doi = {10.1093/mnras/stab1753},
archivePrefix = {arXiv},
       eprint = {2106.09272},
 primaryClass = {astro-ph.HE},
       adsurl = {https://ui.adsabs.harvard.edu/abs/2021MNRAS.506..741M}
}

@ARTICLE{2017ApJ...841...27N,
       author = {{Ni}, Lei and {Zhang}, Qing-Min and {Murphy}, Nicholas A. and {Lin}, Jun},
        title = "{Blob Formation and Ejection in Coronal Jets due to the Plasmoid and Kelvin-Helmholtz Instabilities}",
      journal = {\apj},
         year = 2017,
        month = may,
       volume = {841},
       number = {1},
          eid = {27},
        pages = {27},
          doi = {10.3847/1538-4357/aa6ffe},
archivePrefix = {arXiv},
       eprint = {1705.00180},
 primaryClass = {astro-ph.SR},
       adsurl = {https://ui.adsabs.harvard.edu/abs/2017ApJ...841...27N}
}

@ARTICLE{2011PhPl...18d2105U,
       author = {{Uzdensky}, Dmitri A. and {McKinney}, Jonathan C.},
        title = "{Magnetic reconnection with radiative cooling. I. Optically thin regime}",
      journal = {Physics of Plasmas},
         year = 2011,
        month = apr,
       volume = {18},
       number = {4},
        pages = {042105-042105},
          doi = {10.1063/1.3571602},
archivePrefix = {arXiv},
       eprint = {1007.0774},
 primaryClass = {astro-ph.HE},
       adsurl = {https://ui.adsabs.harvard.edu/abs/2011PhPl...18d2105U}
}

@ARTICLE{1995ApJ...449..777D,
       author = {{Dorman}, Victoria L. and {Kulsrud}, Russell M.},
        title = "{One-dimensional Merging of Magnetic Fields with Cooling}",
      journal = {\apj},
         year = 1995,
        month = aug,
       volume = {449},
        pages = {777},
          doi = {10.1086/176097},
       adsurl = {https://ui.adsabs.harvard.edu/abs/1995ApJ...449..777D}
}

@ARTICLE{2023MNRAS.518..405D,
       author = {{Dihingia}, Indu K. and {Mizuno}, Yosuke and {Fromm}, Christian M. and {Rezzolla}, Luciano},
        title = "{Temperature properties in magnetized and radiatively cooled two-temperature accretion flows on to a black hole}",
      journal = {\mnras},
         year = 2023,
        month = jan,
       volume = {518},
       number = {1},
        pages = {405-417},
          doi = {10.1093/mnras/stac3165},
archivePrefix = {arXiv},
       eprint = {2206.13184},
 primaryClass = {astro-ph.HE},
       adsurl = {https://ui.adsabs.harvard.edu/abs/2023MNRAS.518..405D}
}

@ARTICLE{2012PhPl...19d2303L,
       author = {{Loureiro}, N.~F. and {Samtaney}, R. and {Schekochihin}, A.~A. and {Uzdensky}, D.~A.},
        title = "{Magnetic reconnection and stochastic plasmoid chains in high-Lundquist-number plasmas}",
      journal = {Physics of Plasmas},
         year = 2012,
        month = apr,
       volume = {19},
       number = {4},
        pages = {042303-042303},
          doi = {10.1063/1.3703318},
archivePrefix = {arXiv},
       eprint = {1108.4040},
 primaryClass = {astro-ph.SR},
       adsurl = {https://ui.adsabs.harvard.edu/abs/2012PhPl...19d2303L}
}

@ARTICLE{2020MNRAS.495.1549N,
       author = {{Nathanail}, Antonios and {Fromm}, Christian M. and {Porth}, Oliver and {Olivares}, Hector and {Younsi}, Ziri and {Mizuno}, Yosuke and {Rezzolla}, Luciano},
        title = "{Plasmoid formation in global GRMHD simulations and AGN flares}",
      journal = {\mnras},
         year = 2020,
        month = jun,
       volume = {495},
       number = {2},
        pages = {1549-1565},
          doi = {10.1093/mnras/staa1165},
archivePrefix = {arXiv},
       eprint = {2002.01777},
 primaryClass = {astro-ph.HE},
       adsurl = {https://ui.adsabs.harvard.edu/abs/2020MNRAS.495.1549N}
}

@article{van2014scikit,
  title={scikit-image: image processing in Python},
  author={Van der Walt, Stefan and Sch{\"o}nberger, Johannes L and Nunez-Iglesias, Juan and Boulogne, Fran{\c{c}}ois and Warner, Joshua D and Yager, Neil and Gouillart, Emmanuelle and Yu, Tony},
  journal={PeerJ},
  volume={2},
  pages={e453},
  year={2014},
  publisher={PeerJ Inc.}
}

@ARTICLE{1986PhFl...29.1520B,
       author = {{Biskamp}, D.},
        title = "{Magnetic reconnection via current sheets}",
      journal = {Physics of Fluids},
         year = 1986,
        month = may,
       volume = {29},
       number = {5},
        pages = {1520-1531},
          doi = {10.1063/1.865670},
       adsurl = {https://ui.adsabs.harvard.edu/abs/1986PhFl...29.1520B}
}

@ARTICLE{2020MNRAS.499.3178Y,
       author = {{Yoon}, D. and {Chatterjee}, K. and {Markoff}, S.~B. and {van Eijnatten}, D. and {Younsi}, Z. and {Liska}, M. and {Tchekhovskoy}, A.},
        title = "{Spectral and imaging properties of Sgr A* from high-resolution 3D GRMHD simulations with radiative cooling}",
      journal = {\mnras},
         year = 2020,
        month = dec,
       volume = {499},
       number = {3},
        pages = {3178-3192},
          doi = {10.1093/mnras/staa3031},
archivePrefix = {arXiv},
       eprint = {2009.14227},
 primaryClass = {astro-ph.HE},
       adsurl = {https://ui.adsabs.harvard.edu/abs/2020MNRAS.499.3178Y}
}

@ARTICLE{2022ApJ...930L..12E,
       author = {{Event Horizon Telescope Collaboration} and {Akiyama}, Kazunori and {Alberdi}, Antxon and {Alef}, Walter and {Algaba}, Juan Carlos and {Anantua}, Richard and {Asada}, Keiichi and {Azulay}, Rebecca and {Bach}, Uwe and {Baczko}, Anne-Kathrin and {Ball}, David and {Balokovi{\'c}}, Mislav and {Barrett}, John and {Baub{\"o}ck}, Michi and {Benson}, Bradford A. and {Bintley}, Dan and {Blackburn}, Lindy and {Blundell}, Raymond and {Bouman}, Katherine L. and {Bower}, Geoffrey C. and {Boyce}, Hope and {Bremer}, Michael and {Brinkerink}, Christiaan D. and {Brissenden}, Roger and {Britzen}, Silke and {Broderick}, Avery E. and {Broguiere}, Dominique and {Bronzwaer}, Thomas and {Bustamante}, Sandra and {Byun}, Do-Young and {Carlstrom}, John E. and {Ceccobello}, Chiara and {Chael}, Andrew and {Chan}, Chi-kwan and {Chatterjee}, Koushik and {Chatterjee}, Shami and {Chen}, Ming-Tang and {Chen}, Yongjun and {Cheng}, Xiaopeng and {Cho}, Ilje and {Christian}, Pierre and {Conroy}, Nicholas S. and {Conway}, John E. and {Cordes}, James M. and {Crawford}, Thomas M. and {Crew}, Geoffrey B. and {Cruz-Osorio}, Alejandro and {Cui}, Yuzhu and {Davelaar}, Jordy and {De Laurentis}, Mariafelicia and {Deane}, Roger and {Dempsey}, Jessica and {Desvignes}, Gregory and {Dexter}, Jason and {Dhruv}, Vedant and {Doeleman}, Sheperd S. and {Dougal}, Sean and {Dzib}, Sergio A. and {Eatough}, Ralph P. and {Emami}, Razieh and {Falcke}, Heino and {Farah}, Joseph and {Fish}, Vincent L. and {Fomalont}, Ed and {Ford}, H. Alyson and {Fraga-Encinas}, Raquel and {Freeman}, William T. and {Friberg}, Per and {Fromm}, Christian M. and {Fuentes}, Antonio and {Galison}, Peter and {Gammie}, Charles F. and {Garc{\'\i}a}, Roberto and {Gentaz}, Olivier and {Georgiev}, Boris and {Goddi}, Ciriaco and {Gold}, Roman and {G{\'o}mez-Ruiz}, Arturo I. and {G{\'o}mez}, Jos{\'e} L. and {Gu}, Minfeng and {Gurwell}, Mark and {Hada}, Kazuhiro and {Haggard}, Daryl and {Haworth}, Kari and {Hecht}, Michael H. and {Hesper}, Ronald and {Heumann}, Dirk and {Ho}, Luis C. and {Ho}, Paul and {Honma}, Mareki and {Huang}, Chih-Wei L. and {Huang}, Lei and {Hughes}, David H. and {Ikeda}, Shiro and {Impellizzeri}, C.~M. Violette and {Inoue}, Makoto and {Issaoun}, Sara and {James}, David J. and {Jannuzi}, Buell T. and {Janssen}, Michael and {Jeter}, Britton and {Jiang}, Wu and {Jim{\'e}nez-Rosales}, Alejandra and {Johnson}, Michael D. and {Jorstad}, Svetlana and {Joshi}, Abhishek V. and {Jung}, Taehyun and {Karami}, Mansour and {Karuppusamy}, Ramesh and {Kawashima}, Tomohisa and {Keating}, Garrett K. and {Kettenis}, Mark and {Kim}, Dong-Jin and {Kim}, Jae-Young and {Kim}, Jongsoo and {Kim}, Junhan and {Kino}, Motoki and {Koay}, Jun Yi and {Kocherlakota}, Prashant and {Kofuji}, Yutaro and {Koch}, Patrick M. and {Koyama}, Shoko and {Kramer}, Carsten and {Kramer}, Michael and {Krichbaum}, Thomas P. and {Kuo}, Cheng-Yu and {La Bella}, Noemi and {Lauer}, Tod R. and {Lee}, Daeyoung and {Lee}, Sang-Sung and {Leung}, Po Kin and {Levis}, Aviad and {Li}, Zhiyuan and {Lico}, Rocco and {Lindahl}, Greg and {Lindqvist}, Michael and {Lisakov}, Mikhail and {Liu}, Jun and {Liu}, Kuo and {Liuzzo}, Elisabetta and {Lo}, Wen-Ping and {Lobanov}, Andrei P. and {Loinard}, Laurent and {Lonsdale}, Colin J. and {Lu}, Ru-Sen and {Mao}, Jirong and {Marchili}, Nicola and {Markoff}, Sera and {Marrone}, Daniel P. and {Marscher}, Alan P. and {Mart{\'\i}-Vidal}, Iv{\'a}n and {Matsushita}, Satoki and {Matthews}, Lynn D. and {Medeiros}, Lia and {Menten}, Karl M. and {Michalik}, Daniel and {Mizuno}, Izumi and {Mizuno}, Yosuke and {Moran}, James M. and {Moriyama}, Kotaro and {Moscibrodzka}, Monika and {M{\"u}ller}, Cornelia and {Mus}, Alejandro and {Musoke}, Gibwa and {Myserlis}, Ioannis and {Nadolski}, Andrew and {Nagai}, Hiroshi and {Nagar}, Neil M. and {Nakamura}, Masanori and {Narayan}, Ramesh and {Narayanan}, Gopal and {Natarajan}, Iniyan and {Nathanail}, Antonios and {Fuentes}, Santiago Navarro and {Neilsen}, Joey and {Neri}, Roberto and {Ni}, Chunchong and {Noutsos}, Aristeidis and {Nowak}, Michael A. and {Oh}, Junghwan and {Okino}, Hiroki and {Olivares}, H{\'e}ctor and {Ortiz-Le{\'o}n}, Gisela N. and {Oyama}, Tomoaki and {{\"O}zel}, Feryal and {Palumbo}, Daniel C.~M. and {Paraschos}, Georgios Filippos and {Park}, Jongho and {Parsons}, Harriet and {Patel}, Nimesh and {Pen}, Ue-Li and {Pesce}, Dominic W. and {Pi{\'e}tu}, Vincent and {Plambeck}, Richard and {PopStefanija}, Aleksandar and {Porth}, Oliver and {P{\"o}tzl}, Felix M. and {Prather}, Ben and {Preciado-L{\'o}pez}, Jorge A. and {Psaltis}, Dimitrios and {Pu}, Hung-Yi and {Ramakrishnan}, Venkatessh and {Rao}, Ramprasad and {Rawlings}, Mark G. and {Raymond}, Alexander W. and {Rezzolla}, Luciano and {Ricarte}, Angelo and {Ripperda}, Bart and {Roelofs}, Freek and {Rogers}, Alan and {Ros}, Eduardo and {Romero-Ca{\~n}izales}, Cristina and {Roshanineshat}, Arash and {Rottmann}, Helge and {Roy}, Alan L. and {Ruiz}, Ignacio and {Ruszczyk}, Chet and {Rygl}, Kazi L.~J. and {S{\'a}nchez}, Salvador and {S{\'a}nchez-Arg{\"u}elles}, David and {S{\'a}nchez-Portal}, Miguel and {Sasada}, Mahito and {Satapathy}, Kaushik and {Savolainen}, Tuomas and {Schloerb}, F. Peter and {Schonfeld}, Jonathan and {Schuster}, Karl-Friedrich and {Shao}, Lijing and {Shen}, Zhiqiang and {Small}, Des and {Sohn}, Bong Won and {SooHoo}, Jason and {Souccar}, Kamal and {Sun}, He and {Tazaki}, Fumie and {Tetarenko}, Alexandra J. and {Tiede}, Paul and {Tilanus}, Remo P.~J. and {Titus}, Michael and {Torne}, Pablo and {Traianou}, Efthalia and {Trent}, Tyler and {Trippe}, Sascha and {Turk}, Matthew and {van Bemmel}, Ilse and {van Langevelde}, Huib Jan and {van Rossum}, Daniel R. and {Vos}, Jesse and {Wagner}, Jan and {Ward-Thompson}, Derek and {Wardle}, John and {Weintroub}, Jonathan and {Wex}, Norbert and {Wharton}, Robert and {Wielgus}, Maciek and {Wiik}, Kaj and {Witzel}, Gunther and {Wondrak}, Michael F. and {Wong}, George N. and {Wu}, Qingwen and {Yamaguchi}, Paul and {Yoon}, Doosoo and {Young}, Andr{\'e} and {Young}, Ken and {Younsi}, Ziri and {Yuan}, Feng and {Yuan}, Ye-Fei and {Zensus}, J. Anton and {Zhang}, Shuo and {Zhao}, Guang-Yao and {Zhao}, Shan-Shan and {Agurto}, Claudio and {Allardi}, Alexander and {Amestica}, Rodrigo and {Araneda}, Juan Pablo and {Arriagada}, Oriel and {Berghuis}, Jennie L. and {Bertarini}, Alessandra and {Berthold}, Ryan and {Blanchard}, Jay and {Brown}, Ken and {C{\'a}rdenas}, Mauricio and {Cantzler}, Michael and {Caro}, Patricio and {Castillo-Dom{\'\i}nguez}, Edgar and {Chan}, Tin Lok and {Chang}, Chih-Cheng and {Chang}, Dominic O. and {Chang}, Shu-Hao and {Chang}, Song-Chu and {Chen}, Chung-Chen and {Chilson}, Ryan and {Chuter}, Tim C. and {Ciechanowicz}, Miroslaw and {Colin-Beltran}, Edgar and {Coulson}, Iain M. and {Crowley}, Joseph and {Degenaar}, Nathalie and {Dornbusch}, Sven and {Dur{\'a}n}, Carlos A. and {Everett}, Wendeline B. and {Faber}, Aaron and {Forster}, Karl and {Fuchs}, Miriam M. and {Gale}, David M. and {Geertsema}, Gertie and {Gonz{\'a}lez}, Edouard and {Graham}, Dave and {Gueth}, Fr{\'e}d{\'e}ric and {Halverson}, Nils W. and {Han}, Chih-Chiang and {Han}, Kuo-Chang and {Hasegawa}, Yutaka and {Hern{\'a}ndez-Rebollar}, Jos{\'e} Luis and {Herrera}, Cristian and {Herrero-Illana}, Ruben and {Heyminck}, Stefan and {Hirota}, Akihiko and {Hoge}, James and {Hostler Schimpf}, Shelbi R. and {Howie}, Ryan E. and {Huang}, Yau-De and {Jiang}, Homin and {Jinchi}, Hao and {John}, David and {Kimura}, Kimihiro and {Klein}, Thomas and {Kubo}, Derek and {Kuroda}, John and {Kwon}, Caleb and {Lacasse}, Richard and {Laing}, Robert and {Leitch}, Erik M. and {Li}, Chao-Te and {Liu}, Ching-Tang and {Liu}, Kuan-Yu and {Lin}, Lupin C. -C. and {Lu}, Li-Ming and {Mac-Auliffe}, Felipe and {Martin-Cocher}, Pierre and {Matulonis}, Callie and {Maute}, John K. and {Messias}, Hugo and {Meyer-Zhao}, Zheng and {Monta{\~n}a}, Alfredo and {Montenegro-Montes}, Francisco and {Montgomerie}, William and {Moreno Nolasco}, Marcos Emir and {Muders}, Dirk and {Nishioka}, Hiroaki and {Norton}, Timothy J. and {Nystrom}, George and {Ogawa}, Hideo and {Olivares}, Rodrigo and {Oshiro}, Peter and {P{\'e}rez-Beaupuits}, Juan Pablo and {Parra}, Rodrigo and {Phillips}, Neil M. and {Poirier}, Michael and {Pradel}, Nicolas and {Qiu}, Richard and {Raffin}, Philippe A. and {Rahlin}, Alexandra S. and {Ram{\'\i}rez}, Jorge and {Ressler}, Sean and {Reynolds}, Mark and {Rodr{\'\i}guez-Montoya}, Iv{\'a}n and {Saez-Madain}, Alejandro F. and {Santana}, Jorge and {Shaw}, Paul and {Shirkey}, Leslie E. and {Silva}, Kevin M. and {Snow}, William and {Sousa}, Don and {Sridharan}, T.~K. and {Stahm}, William and {Stark}, Anthony A. and {Test}, John and {Torstensson}, Karl and {Venegas}, Paulina and {Walther}, Craig and {Wei}, Ta-Shun and {White}, Chris and {Wieching}, Gundolf and {Wijnands}, Rudy and {Wouterloot}, Jan G.~A. and {Yu}, Chen-Yu and {Yu (于威)}, Wei and {Zeballos}, Milagros},
        title = "{First Sagittarius A* Event Horizon Telescope Results. I. The Shadow of the Supermassive Black Hole in the Center of the Milky Way}",
      journal = {\apjl},
         year = 2022,
        month = may,
       volume = {930},
       number = {2},
          eid = {L12},
        pages = {L12},
          doi = {10.3847/2041-8213/ac6674},
       adsurl = {https://ui.adsabs.harvard.edu/abs/2022ApJ...930L..12E}
}

@ARTICLE{2022MNRAS.513.4267N,
       author = {{Nathanail}, Antonios and {Mpisketzis}, Vasilis and {Porth}, Oliver and {Fromm}, Christian M. and {Rezzolla}, Luciano},
        title = "{Magnetic reconnection and plasmoid formation in three-dimensional accretion flows around black holes}",
      journal = {\mnras},
         year = 2022,
        month = jul,
       volume = {513},
       number = {3},
        pages = {4267-4277},
          doi = {10.1093/mnras/stac1118},
archivePrefix = {arXiv},
       eprint = {2111.03689},
 primaryClass = {astro-ph.HE},
       adsurl = {https://ui.adsabs.harvard.edu/abs/2022MNRAS.513.4267N}
}

@ARTICLE{2016MNRAS.462.3325P,
       author = {{Petropoulou}, Maria and {Giannios}, Dimitrios and {Sironi}, Lorenzo},
        title = "{Blazar flares powered by plasmoids in relativistic reconnection}",
      journal = {\mnras},
         year = 2016,
        month = nov,
       volume = {462},
       number = {3},
        pages = {3325-3343},
          doi = {10.1093/mnras/stw1832},
archivePrefix = {arXiv},
       eprint = {1606.07447},
 primaryClass = {astro-ph.HE},
       adsurl = {https://ui.adsabs.harvard.edu/abs/2016MNRAS.462.3325P}
}

@ARTICLE{2018ApJ...862L..25Z,
       author = {{Zhang}, Haocheng and {Li}, Xiaocan and {Guo}, Fan and {Giannios}, Dimitrios},
        title = "{Large-amplitude Blazar Polarization Angle Swing as a Signature of Magnetic Reconnection}",
      journal = {\apjl},
         year = 2018,
        month = aug,
       volume = {862},
       number = {2},
          eid = {L25},
        pages = {L25},
          doi = {10.3847/2041-8213/aad54f},
archivePrefix = {arXiv},
       eprint = {1807.08420},
 primaryClass = {astro-ph.HE},
       adsurl = {https://ui.adsabs.harvard.edu/abs/2018ApJ...862L..25Z}
}

@ARTICLE{2019MNRAS.482...65C,
       author = {{Christie}, I.~M. and {Petropoulou}, M. and {Sironi}, L. and {Giannios}, D.},
        title = "{Radiative signatures of plasmoid-dominated reconnection in blazar jets}",
      journal = {\mnras},
         year = 2019,
        month = jan,
       volume = {482},
       number = {1},
        pages = {65-82},
          doi = {10.1093/mnras/sty2636},
archivePrefix = {arXiv},
       eprint = {1807.08041},
 primaryClass = {astro-ph.HE},
       adsurl = {https://ui.adsabs.harvard.edu/abs/2019MNRAS.482...65C}
}

@ARTICLE{2012MNRAS.426.1928D,
       author = {{Dibi}, S. and {Drappeau}, S. and {Fragile}, P.~C. and {Markoff}, S. and {Dexter}, J.},
        title = "{General relativistic magnetohydrodynamic simulations of accretion on to Sgr A*: how important are radiative losses?}",
      journal = {\mnras},
         year = 2012,
        month = nov,
       volume = {426},
       number = {3},
        pages = {1928-1939},
          doi = {10.1111/j.1365-2966.2012.21857.x},
archivePrefix = {arXiv},
       eprint = {1206.3976},
 primaryClass = {astro-ph.HE},
       adsurl = {https://ui.adsabs.harvard.edu/abs/2012MNRAS.426.1928D}
}

@ARTICLE{2013MNRAS.431.2872D,
       author = {{Drappeau}, S. and {Dibi}, S. and {Dexter}, J. and {Markoff}, S. and {Fragile}, P.~C.},
        title = "{Self-consistent spectra from radiative GRMHD simulations of accretion on to Sgr A*}",
      journal = {\mnras},
         year = 2013,
        month = may,
       volume = {431},
       number = {3},
        pages = {2872-2884},
          doi = {10.1093/mnras/stt388},
archivePrefix = {arXiv},
       eprint = {1209.4599},
 primaryClass = {astro-ph.HE},
       adsurl = {https://ui.adsabs.harvard.edu/abs/2013MNRAS.431.2872D}
}

@ARTICLE{2009ApJ...693..771F,
       author = {{Fragile}, P. Chris and {Meier}, David L.},
        title = "{General Relativistic Magnetohydrodynamic Simulations of the Hard State as a Magnetically Dominated Accretion Flow}",
      journal = {\apj},
         year = 2009,
        month = mar,
       volume = {693},
       number = {1},
        pages = {771-783},
          doi = {10.1088/0004-637X/693/1/771},
archivePrefix = {arXiv},
       eprint = {0810.1082},
 primaryClass = {astro-ph},
       adsurl = {https://ui.adsabs.harvard.edu/abs/2009ApJ...693..771F}
}

@ARTICLE{2007A&A...470..475A,
       author = {{Aharonian}, F. and {Akhperjanian}, A.~G. and {Bazer-Bachi}, A.~R. and {Beilicke}, M. and {Benbow}, W. and {Berge}, D. and {Bernl{\"o}hr}, K. and {Boisson}, C. and {Bolz}, O. and {Borrel}, V. and {Braun}, I. and {Brion}, E. and {Brown}, A.~M. and {B{\"u}hler}, R. and {B{\"u}sching}, I. and {Boutelier}, T. and {Carrigan}, S. and {Chadwick}, P.~M. and {Chounet}, L. -M. and {Coignet}, G. and {Cornils}, R. and {Costamante}, L. and {Degrange}, B. and {Dickinson}, H.~J. and {Djannati-Ata{\"\i}}, A. and {O'C. Drury}, L. and {Dubus}, G. and {Egberts}, K. and {Emmanoulopoulos}, D. and {Espigat}, P. and {Farnier}, C. and {Feinstein}, F. and {Ferrero}, E. and {Fiasson}, A. and {Fontaine}, G. and {Funk}, Seb. and {Funk}, S. and {F{\"u}{\ss}ling}, M. and {Gallant}, Y.~A. and {Giebels}, B. and {Glicenstein}, J.~F. and {Gl{\"u}ck}, B. and {Goret}, P. and {Hadjichristidis}, C. and {Hauser}, D. and {Hauser}, M. and {Heinzelmann}, G. and {Henri}, G. and {Hermann}, G. and {Hinton}, J.~A. and {Hoffmann}, A. and {Hofmann}, W. and {Holleran}, M. and {Hoppe}, S. and {Horns}, D. and {Jacholkowska}, A. and {de Jager}, O.~C. and {Kendziorra}, E. and {Kerschhaggl}, M. and {Kh{\'e}lifi}, B. and {Komin}, Nu. and {Kosack}, K. and {Lamanna}, G. and {Latham}, I.~J. and {Le Gallou}, R. and {Lemi{\`e}re}, A. and {Lemoine-Goumard}, M. and {Lohse}, T. and {Martin}, J.~M. and {Martineau-Huynh}, O. and {Marcowith}, A. and {Masterson}, C. and {Maurin}, G. and {McComb}, T.~J.~L. and {Moulin}, E. and {de Naurois}, M. and {Nedbal}, D. and {Nolan}, S.~J. and {Noutsos}, A. and {Olive}, J. -P. and {Orford}, K.~J. and {Osborne}, J.~L. and {Panter}, M. and {Pelletier}, G. and {Petrucci}, P. -O. and {Pita}, S. and {P{\"u}hlhofer}, G. and {Punch}, M. and {Ranchon}, S. and {Raubenheimer}, B.~C. and {Raue}, M. and {Rayner}, S.~M. and {Ripken}, J. and {Rob}, L. and {Rolland}, L. and {Rosier-Lees}, S. and {Rowell}, G. and {Sahakian}, V. and {Santangelo}, A. and {Saug{\'e}}, L. and {Schlenker}, S. and {Schlickeiser}, R. and {Schr{\"o}der}, R. and {Schwanke}, U. and {Schwarzburg}, S. and {Schwemmer}, S. and {Shalchi}, A. and {Sol}, H. and {Spangler}, D. and {Spanier}, F. and {Steenkamp}, R. and {Stegmann}, C. and {Superina}, G. and {Tam}, P.~H. and {Tavernet}, J. -P. and {Terrier}, R. and {Tluczykont}, M. and {van Eldik}, C. and {Vasileiadis}, G. and {Venter}, C. and {Vialle}, J.~P. and {Vincent}, P. and {V{\"o}lk}, H.~J. and {Wagner}, S.~J. and {Ward}, M.},
        title = "{Detection of VHE gamma-ray emission from the distant blazar 1ES{\,}1101-232 with HESS and broadband characterisation}",
      journal = {\aap},
         year = 2007,
        month = aug,
       volume = {470},
       number = {2},
        pages = {475-489},
          doi = {10.1051/0004-6361:20077057},
archivePrefix = {arXiv},
       eprint = {0705.2946},
 primaryClass = {astro-ph},
       adsurl = {https://ui.adsabs.harvard.edu/abs/2007A&A...470..475A}
}

@ARTICLE{2007ApJ...663..125A,
       author = {{Albert}, J. and {Aliu}, E. and {Anderhub}, H. and {Antoranz}, P. and {Armada}, A. and {Asensio}, M. and {Baixeras}, C. and {Barrio}, J.~A. and {Bartko}, H. and {Bastieri}, D. and {Becker}, J. and {Bednarek}, W. and {Berger}, K. and {Bigongiari}, C. and {Biland}, A. and {Bock}, R.~K. and {Bordas}, P. and {Bosch-Ramon}, V. and {Bretz}, T. and {Britvitch}, I. and {Camara}, M. and {Carmona}, E. and {Chilingarian}, A. and {Ciprini}, S. and {Coarasa}, J.~A. and {Commichau}, S. and {Contreras}, J.~L. and {Cortina}, J. and {Curtef}, V. and {Danielyan}, V. and {Dazzi}, F. and {De Angelis}, A. and {de los Reyes}, R. and {De Lotto}, B. and {Domingo-Santamar{\'\i}a}, E. and {Dorner}, D. and {Doro}, M. and {Errando}, M. and {Fagiolini}, M. and {Ferenc}, D. and {Fern{\'a}ndez}, E. and {Firpo}, R. and {Flix}, J. and {Fonseca}, M.~V. and {Font}, L. and {Fuchs}, M. and {Galante}, N. and {Garczarczyk}, M. and {Gaug}, M. and {Giller}, M. and {Goebel}, F. and {Hakobyan}, D. and {Hayashida}, M. and {Hengstebeck}, T. and {H{\"o}hne}, D. and {Hose}, J. and {Hsu}, C.~C. and {Jacon}, P. and {Jogler}, T. and {Kalekin}, O. and {Kosyra}, R. and {Kranich}, D. and {Kritzer}, R. and {Laatiaoui}, M. and {Laille}, A. and {Liebing}, P. and {Lindfors}, E. and {Lombardi}, S. and {Longo}, F. and {L{\'o}pez}, J. and {L{\'o}pez}, M. and {Lorenz}, E. and {Majumdar}, P. and {Maneva}, G. and {Mannheim}, K. and {Mansutti}, O. and {Mariotti}, M. and {Mart{\'\i}nez}, M. and {Mazin}, D. and {Merck}, C. and {Meucci}, M. and {Meyer}, M. and {Miranda}, J.~M. and {Mirzoyan}, R. and {Mizobuchi}, S. and {Moralejo}, A. and {Nilsson}, K. and {Ninkovic}, J. and {O{\~n}a-Wilhelmi}, E. and {Ordu{\~n}a}, R. and {Otte}, N. and {Oya}, I. and {Paneque}, D. and {Paoletti}, R. and {Paredes}, J.~M. and {Pasanen}, M. and {Pascoli}, D. and {Pauss}, F. and {Pegna}, R. and {Persic}, M. and {Peruzzo}, L. and {Piccioli}, A. and {Poller}, M. and {Prandini}, E. and {Raymers}, A. and {Rhode}, W. and {Rib{\'o}}, M. and {Rico}, J. and {Rissi}, M. and {Robert}, A. and {R{\"u}gamer}, S. and {Saggion}, A. and {S{\'a}nchez}, A. and {Sartori}, P. and {Scalzotto}, V. and {Scapin}, V. and {Schmitt}, R. and {Schweizer}, T. and {Shayduk}, M. and {Shinozaki}, K. and {Shore}, S.~N. and {Sidro}, N. and {Sillanp{\"a}{\"a}}, A. and {Sobczynska}, D. and {Stamerra}, A. and {Stark}, L.~S. and {Takalo}, L. and {Temnikov}, P. and {Tescaro}, D. and {Teshima}, M. and {Tonello}, N. and {Torres}, A. and {Torres}, D.~F. and {Turini}, N. and {Vankov}, H. and {Vitale}, V. and {Wagner}, R.~M. and {Wibig}, T. and {Wittek}, W. and {Zanin}, R. and {Zapatero}, J.},
        title = "{Observations of Markarian 421 with the MAGIC Telescope}",
      journal = {\apj},
         year = 2007,
        month = jul,
       volume = {663},
       number = {1},
        pages = {125-138},
          doi = {10.1086/518221},
archivePrefix = {arXiv},
       eprint = {astro-ph/0603478},
 primaryClass = {astro-ph},
       adsurl = {https://ui.adsabs.harvard.edu/abs/2007ApJ...663..125A}
}

@ARTICLE{1965ApJ...142..531F,
       author = {{Field}, George B.},
        title = "{Thermal Instability.}",
      journal = {\apj},
         year = 1965,
        month = aug,
       volume = {142},
        pages = {531},
          doi = {10.1086/148317},
       adsurl = {https://ui.adsabs.harvard.edu/abs/1965ApJ...142..531F}
}

@ARTICLE{1991SoPh..135..361F,
       author = {{Forbes}, T.~G. and {Malherbe}, J.~M.},
        title = "{A numerical simulation of magnetic reconnection and radiative cooling in line-tied current sheets}",
      journal = {\solphys},
         year = 1991,
        month = oct,
       volume = {135},
       number = {2},
        pages = {361-391},
          doi = {10.1007/BF00147508},
       adsurl = {https://ui.adsabs.harvard.edu/abs/1991SoPh..135..361F}
}

@ARTICLE{2007A&A...473...11D,
       author = {{Del Zanna}, L. and {Zanotti}, O. and {Bucciantini}, N. and {Londrillo}, P.},
        title = "{ECHO: a Eulerian conservative high-order scheme for general relativistic magnetohydrodynamics and magnetodynamics}",
      journal = {\aap},
         year = 2007,
        month = oct,
       volume = {473},
       number = {1},
        pages = {11-30},
          doi = {10.1051/0004-6361:20077093},
archivePrefix = {arXiv},
       eprint = {0704.3206},
 primaryClass = {astro-ph},
       adsurl = {https://ui.adsabs.harvard.edu/abs/2007A&A...473...11D}
}

@ARTICLE{2024A&A...688A..82J,
       author = {{Jiang}, Hong-Xuan and {Mizuno}, Yosuke and {Dihingia}, Indu K. and {Nathanail}, Antonios and {Younsi}, Ziri and {Fromm}, Christian M.},
        title = "{Dynamics and emission properties of flux ropes from two-temperature GRMHD simulations with multiple magnetic loops}",
      journal = {\aap},
         year = 2024,
        month = aug,
       volume = {688},
          eid = {A82},
        pages = {A82},
          doi = {10.1051/0004-6361/202449681},
archivePrefix = {arXiv},
       eprint = {2404.03237},
 primaryClass = {astro-ph.HE},
       adsurl = {https://ui.adsabs.harvard.edu/abs/2024A&A...688A..82J}
}

@ARTICLE{2004ApJ...606..894Y,
       author = {{Yuan}, Feng and {Quataert}, Eliot and {Narayan}, Ramesh},
        title = "{On the Nature of the Variable Infrared Emission from Sagittarius A*}",
      journal = {\apj},
         year = 2004,
        month = may,
       volume = {606},
       number = {2},
        pages = {894-899},
          doi = {10.1086/383117},
archivePrefix = {arXiv},
       eprint = {astro-ph/0401429},
 primaryClass = {astro-ph},
       adsurl = {https://ui.adsabs.harvard.edu/abs/2004ApJ...606..894Y}
}

@ARTICLE{2007MNRAS.375..764T,
       author = {{Trippe}, S. and {Paumard}, T. and {Ott}, T. and {Gillessen}, S. and {Eisenhauer}, F. and {Martins}, F. and {Genzel}, R.},
        title = "{A polarized infrared flare from Sagittarius A* and the signatures of orbiting plasma hotspots}",
      journal = {\mnras},
         year = 2007,
        month = mar,
       volume = {375},
       number = {3},
        pages = {764-772},
          doi = {10.1111/j.1365-2966.2006.11338.x},
archivePrefix = {arXiv},
       eprint = {astro-ph/0611737},
 primaryClass = {astro-ph},
       adsurl = {https://ui.adsabs.harvard.edu/abs/2007MNRAS.375..764T}
}

@INPROCEEDINGS{2022AAS...24021101M,
       author = {{Markoff}, Sera and {Event Horizon Telescope Collaboration}},
        title = "{First Sagittarius A* Event Horizon Telescope Results: The Shadow of the Supermassive Black Hole in the Center of the Milky Way}",
    booktitle = {American Astronomical Society Meeting \#240},
         year = 2022,
       series = {American Astronomical Society Meeting Abstracts},
       volume = {240},
        month = jun,
          eid = {211.01},
        pages = {211.01},
       adsurl = {https://ui.adsabs.harvard.edu/abs/2022AAS...24021101M}
}

@ARTICLE{2022ApJ...935L...1L,
       author = {{Liska}, M.~T.~P. and {Musoke}, G. and {Tchekhovskoy}, A. and {Porth}, O. and {Beloborodov}, A.~M.},
        title = "{Formation of Magnetically Truncated Accretion Disks in 3D Radiation-transport Two-temperature GRMHD Simulations}",
      journal = {\apjl},
         year = 2022,
        month = aug,
       volume = {935},
       number = {1},
          eid = {L1},
        pages = {L1},
          doi = {10.3847/2041-8213/ac84db},
archivePrefix = {arXiv},
       eprint = {2201.03526},
 primaryClass = {astro-ph.HE},
       adsurl = {https://ui.adsabs.harvard.edu/abs/2022ApJ...935L...1L}
}

@ARTICLE{2009ApJ...699.1789T,
       author = {{Tchekhovskoy}, Alexander and {McKinney}, Jonathan C. and {Narayan}, Ramesh},
        title = "{Efficiency of Magnetic to Kinetic Energy Conversion in a Monopole Magnetosphere}",
      journal = {\apj},
         year = 2009,
        month = jul,
       volume = {699},
       number = {2},
        pages = {1789-1808},
          doi = {10.1088/0004-637X/699/2/1789},
archivePrefix = {arXiv},
       eprint = {0901.4776},
 primaryClass = {astro-ph.HE},
       adsurl = {https://ui.adsabs.harvard.edu/abs/2009ApJ...699.1789T}
}

@ARTICLE{2024MNRAS.530.1563G,
       author = {{Grigorian}, A.~A. and {Dexter}, J.},
        title = "{The relationship between simulated sub-millimeter and near-infrared images of Sagittarius A* from a magnetically arrested black hole accretion flow}",
      journal = {\mnras},
         year = 2024,
        month = may,
       volume = {530},
       number = {2},
        pages = {1563-1579},
          doi = {10.1093/mnras/stae934},
archivePrefix = {arXiv},
       eprint = {2404.10982},
 primaryClass = {astro-ph.HE},
       adsurl = {https://ui.adsabs.harvard.edu/abs/2024MNRAS.530.1563G}
}

@ARTICLE{2022MNRAS.511.3536S,
       author = {{Scepi}, Nicolas and {Dexter}, Jason and {Begelman}, Mitchell C.},
        title = "{Sgr A* X-ray flares from non-thermal particle acceleration in a magnetically arrested disc}",
      journal = {\mnras},
         year = 2022,
        month = apr,
       volume = {511},
       number = {3},
        pages = {3536-3547},
          doi = {10.1093/mnras/stac337},
archivePrefix = {arXiv},
       eprint = {2107.08056},
 primaryClass = {astro-ph.HE},
       adsurl = {https://ui.adsabs.harvard.edu/abs/2022MNRAS.511.3536S}
}

@ARTICLE{2025ApJ...981L..11S,
       author = {{Singh}, Akshay and {B{\'e}gu{\'e}}, Damien and {Pe'er}, Asaf},
        title = "{Radiative Cooling Changes the Dynamics of Magnetically Arrested Disks}",
      journal = {\apjl},
         year = 2025,
        month = mar,
       volume = {981},
       number = {1},
          eid = {L11},
        pages = {L11},
          doi = {10.3847/2041-8213/adb749},
archivePrefix = {arXiv},
       eprint = {2412.11440},
 primaryClass = {astro-ph.HE},
       adsurl = {https://ui.adsabs.harvard.edu/abs/2025ApJ...981L..11S}
}

@ARTICLE{2022ApJ...930L..16E,
       author = {{Event Horizon Telescope Collaboration} and {Akiyama}, Kazunori and {Alberdi}, Antxon and {Alef}, Walter and {Algaba}, Juan Carlos and {Anantua}, Richard and {Asada}, Keiichi and {Azulay}, Rebecca and {Bach}, Uwe and {Baczko}, Anne-Kathrin and {Ball}, David and {Balokovi{\'c}}, Mislav and {Barrett}, John and {Baub{\"o}ck}, Michi and {Benson}, Bradford A. and {Bintley}, Dan and {Blackburn}, Lindy and {Blundell}, Raymond and {Bouman}, Katherine L. and {Bower}, Geoffrey C. and {Boyce}, Hope and {Bremer}, Michael and {Brinkerink}, Christiaan D. and {Brissenden}, Roger and {Britzen}, Silke and {Broderick}, Avery E. and {Broguiere}, Dominique and {Bronzwaer}, Thomas and {Bustamante}, Sandra and {Byun}, Do-Young and {Carlstrom}, John E. and {Ceccobello}, Chiara and {Chael}, Andrew and {Chan}, Chi-kwan and {Chatterjee}, Koushik and {Chatterjee}, Shami and {Chen}, Ming-Tang and {Chen}, Yongjun and {Cheng}, Xiaopeng and {Cho}, Ilje and {Christian}, Pierre and {Conroy}, Nicholas S. and {Conway}, John E. and {Cordes}, James M. and {Crawford}, Thomas M. and {Crew}, Geoffrey B. and {Cruz-Osorio}, Alejandro and {Cui}, Yuzhu and {Davelaar}, Jordy and {De Laurentis}, Mariafelicia and {Deane}, Roger and {Dempsey}, Jessica and {Desvignes}, Gregory and {Dexter}, Jason and {Dhruv}, Vedant and {Doeleman}, Sheperd S. and {Dougal}, Sean and {Dzib}, Sergio A. and {Eatough}, Ralph P. and {Emami}, Razieh and {Falcke}, Heino and {Farah}, Joseph and {Fish}, Vincent L. and {Fomalont}, Ed and {Ford}, H. Alyson and {Fraga-Encinas}, Raquel and {Freeman}, William T. and {Friberg}, Per and {Fromm}, Christian M. and {Fuentes}, Antonio and {Galison}, Peter and {Gammie}, Charles F. and {Garc{\'\i}a}, Roberto and {Gentaz}, Olivier and {Georgiev}, Boris and {Goddi}, Ciriaco and {Gold}, Roman and {G{\'o}mez-Ruiz}, Arturo I. and {G{\'o}mez}, Jos{\'e} L. and {Gu}, Minfeng and {Gurwell}, Mark and {Hada}, Kazuhiro and {Haggard}, Daryl and {Haworth}, Kari and {Hecht}, Michael H. and {Hesper}, Ronald and {Heumann}, Dirk and {Ho}, Luis C. and {Ho}, Paul and {Honma}, Mareki and {Huang}, Chih-Wei L. and {Huang}, Lei and {Hughes}, David H. and {Ikeda}, Shiro and {Violette Impellizzeri}, C.~M. and {Inoue}, Makoto and {Issaoun}, Sara and {James}, David J. and {Jannuzi}, Buell T. and {Janssen}, Michael and {Jeter}, Britton and {Jiang}, Wu and {Jim{\'e}nez-Rosales}, Alejandra and {Johnson}, Michael D. and {Jorstad}, Svetlana and {Joshi}, Abhishek V. and {Jung}, Taehyun and {Karami}, Mansour and {Karuppusamy}, Ramesh and {Kawashima}, Tomohisa and {Keating}, Garrett K. and {Kettenis}, Mark and {Kim}, Dong-Jin and {Kim}, Jae-Young and {Kim}, Jongsoo and {Kim}, Junhan and {Kino}, Motoki and {Koay}, Jun Yi and {Kocherlakota}, Prashant and {Kofuji}, Yutaro and {Koch}, Patrick M. and {Koyama}, Shoko and {Kramer}, Carsten and {Kramer}, Michael and {Krichbaum}, Thomas P. and {Kuo}, Cheng-Yu and {La Bella}, Noemi and {Lauer}, Tod R. and {Lee}, Daeyoung and {Lee}, Sang-Sung and {Leung}, Po Kin and {Levis}, Aviad and {Li}, Zhiyuan and {Lico}, Rocco and {Lindahl}, Greg and {Lindqvist}, Michael and {Lisakov}, Mikhail and {Liu}, Jun and {Liu}, Kuo and {Liuzzo}, Elisabetta and {Lo}, Wen-Ping and {Lobanov}, Andrei P. and {Loinard}, Laurent and {Lonsdale}, Colin J. and {Lu}, Ru-Sen and {Mao}, Jirong and {Marchili}, Nicola and {Markoff}, Sera and {Marrone}, Daniel P. and {Marscher}, Alan P. and {Mart{\'\i}-Vidal}, Iv{\'a}n and {Matsushita}, Satoki and {Matthews}, Lynn D. and {Medeiros}, Lia and {Menten}, Karl M. and {Michalik}, Daniel and {Mizuno}, Izumi and {Mizuno}, Yosuke and {Moran}, James M. and {Moriyama}, Kotaro and {Moscibrodzka}, Monika and {M{\"u}ller}, Cornelia and {Mus}, Alejandro and {Musoke}, Gibwa and {Myserlis}, Ioannis and {Nadolski}, Andrew and {Nagai}, Hiroshi and {Nagar}, Neil M. and {Nakamura}, Masanori and {Narayan}, Ramesh and {Narayanan}, Gopal and {Natarajan}, Iniyan and {Nathanail}, Antonios and {Navarro Fuentes}, Santiago and {Neilsen}, Joey and {Neri}, Roberto and {Ni}, Chunchong and {Noutsos}, Aristeidis and {Nowak}, Michael A. and {Oh}, Junghwan and {Okino}, Hiroki and {Olivares}, H{\'e}ctor and {Ortiz-Le{\'o}n}, Gisela N. and {Oyama}, Tomoaki and {{\"O}zel}, Feryal and {Palumbo}, Daniel C.~M. and {Filippos Paraschos}, Georgios and {Park}, Jongho and {Parsons}, Harriet and {Patel}, Nimesh and {Pen}, Ue-Li and {Pesce}, Dominic W. and {Pi{\'e}tu}, Vincent and {Plambeck}, Richard and {PopStefanija}, Aleksandar and {Porth}, Oliver and {P{\"o}tzl}, Felix M. and {Prather}, Ben and {Preciado-L{\'o}pez}, Jorge A. and {Psaltis}, Dimitrios},
        title = "{First Sagittarius A* Event Horizon Telescope Results. V. Testing Astrophysical Models of the Galactic Center Black Hole}",
      journal = {\apjl},
         year = 2022,
        month = may,
       volume = {930},
       number = {2},
          eid = {L16},
        pages = {L16},
          doi = {10.3847/2041-8213/ac6672},
       adsurl = {https://ui.adsabs.harvard.edu/abs/2022ApJ...930L..16E}
}

@ARTICLE{2018ApJ...864..126R,
       author = {{Ryan}, Benjamin R. and {Ressler}, Sean M. and {Dolence}, Joshua C. and {Gammie}, Charles and {Quataert}, Eliot},
        title = "{Two-temperature GRRMHD Simulations of M87}",
      journal = {\apj},
         year = 2018,
        month = sep,
       volume = {864},
       number = {2},
          eid = {126},
        pages = {126},
          doi = {10.3847/1538-4357/aad73a},
archivePrefix = {arXiv},
       eprint = {1808.01958},
 primaryClass = {astro-ph.HE},
       adsurl = {https://ui.adsabs.harvard.edu/abs/2018ApJ...864..126R}
}

@ARTICLE{2025MNRAS.537.2496C,
       author = {{Chael}, Andrew},
        title = "{Survey of radiative, two-temperature magnetically arrested simulations of the black hole M87* I: turbulent electron heating}",
      journal = {\mnras},
         year = 2025,
        month = mar,
       volume = {537},
       number = {3},
        pages = {2496-2515},
          doi = {10.1093/mnras/staf200},
archivePrefix = {arXiv},
       eprint = {2501.12448},
 primaryClass = {astro-ph.HE},
       adsurl = {https://ui.adsabs.harvard.edu/abs/2025MNRAS.537.2496C}
}

@ARTICLE{2025MNRAS.538..698S,
       author = {{Salas}, L.~D.~S. and {Liska}, M.~T.~P. and {Markoff}, S.~B. and {Chatterjee}, K. and {Musoke}, G. and {Porth}, O. and {Ripperda}, B. and {Yoon}, D. and {Mulaudzi}, W.},
        title = "{Two-temperature treatments in magnetically arrested disc GRMHD simulations more accurately predict light curves of Sagittarius A*}",
      journal = {\mnras},
         year = 2025,
        month = apr,
       volume = {538},
       number = {2},
        pages = {698-710},
          doi = {10.1093/mnras/staf240},
archivePrefix = {arXiv},
       eprint = {2411.09556},
 primaryClass = {astro-ph.HE},
       adsurl = {https://ui.adsabs.harvard.edu/abs/2025MNRAS.538..698S}
}

@ARTICLE{2014PhRvD..89b4041L,
       author = {{Lasota}, J. -P. and {Gourgoulhon}, E. and {Abramowicz}, M. and {Tchekhovskoy}, A. and {Narayan}, R.},
        title = "{Extracting black-hole rotational energy: The generalized Penrose process}",
      journal = {\prd},
         year = 2014,
        month = jan,
       volume = {89},
       number = {2},
          eid = {024041},
        pages = {024041},
          doi = {10.1103/PhysRevD.89.024041},
archivePrefix = {arXiv},
       eprint = {1310.7499},
 primaryClass = {gr-qc},
       adsurl = {https://ui.adsabs.harvard.edu/abs/2014PhRvD..89b4041L}
}

@ARTICLE{2015PhRvL.114k5003A,
       author = {{Asenjo}, Felipe A. and {Comisso}, Luca},
        title = "{Generalized Magnetofluid Connections in Relativistic Magnetohydrodynamics}",
      journal = {\prl},
         year = 2015,
        month = mar,
       volume = {114},
       number = {11},
          eid = {115003},
        pages = {115003},
          doi = {10.1103/PhysRevLett.114.115003},
archivePrefix = {arXiv},
       eprint = {1502.07461},
 primaryClass = {physics.plasm-ph},
       adsurl = {https://ui.adsabs.harvard.edu/abs/2015PhRvL.114k5003A}
}

@ARTICLE{2025PhRvD.111b3003S,
       author = {{Shen}, Ye and {YuChih}, Ho-Yun},
        title = "{Energy extraction from a rotating black hole via magnetic reconnection: Parameters in reconnection models}",
      journal = {\prd},
         year = 2025,
        month = jan,
       volume = {111},
       number = {2},
          eid = {023003},
        pages = {023003},
          doi = {10.1103/PhysRevD.111.023003},
archivePrefix = {arXiv},
       eprint = {2412.03010},
 primaryClass = {astro-ph.HE},
       adsurl = {https://ui.adsabs.harvard.edu/abs/2025PhRvD.111b3003S}
}

@ARTICLE{2024PhRvD.110j4044F,
       author = {{Fan}, Zhong-Ying and {Li}, Yuehang and {Zhou}, Fan and {Guo}, Minyong},
        title = "{Fast magnetic reconnection in Kerr spacetime}",
      journal = {\prd},
         year = 2024,
        month = nov,
       volume = {110},
       number = {10},
          eid = {104044},
        pages = {104044},
          doi = {10.1103/PhysRevD.110.104044},
archivePrefix = {arXiv},
       eprint = {2409.05434},
 primaryClass = {astro-ph.HE},
       adsurl = {https://ui.adsabs.harvard.edu/abs/2024PhRvD.110j4044F}
}

@ARTICLE{2008ApJ...682.1124K,
       author = {{Koide}, Shinji and {Arai}, Kenzo},
        title = "{Energy Extraction from a Rotating Black Hole by Magnetic Reconnection in the Ergosphere}",
      journal = {\apj},
         year = 2008,
        month = aug,
       volume = {682},
       number = {2},
        pages = {1124-1133},
          doi = {10.1086/589497},
archivePrefix = {arXiv},
       eprint = {0805.0044},
 primaryClass = {astro-ph},
       adsurl = {https://ui.adsabs.harvard.edu/abs/2008ApJ...682.1124K}
}

@ARTICLE{2025ApJ...982L..31C,
       author = {{Camilloni}, Filippo and {Rezzolla}, Luciano},
        title = "{Self-consistent Multidimensional Penrose Process Driven by Magnetic Reconnection}",
      journal = {\apjl},
         year = 2025,
        month = mar,
       volume = {982},
       number = {1},
          eid = {L31},
        pages = {L31},
          doi = {10.3847/2041-8213/adbbef},
archivePrefix = {arXiv},
       eprint = {2411.04184},
 primaryClass = {gr-qc},
       adsurl = {https://ui.adsabs.harvard.edu/abs/2025ApJ...982L..31C}
}

@ARTICLE{2025A&A...696A..36D,
       author = {{Dimitropoulos}, I. and {Nathanail}, A. and {Petropoulou}, M. and {Contopoulos}, I. and {Fromm}, C.~M.},
        title = "{Flares from plasmoids and current sheets around Sgr A*}",
      journal = {\aap},
         year = 2025,
        month = apr,
       volume = {696},
          eid = {A36},
        pages = {A36},
          doi = {10.1051/0004-6361/202451577},
archivePrefix = {arXiv},
       eprint = {2407.14312},
 primaryClass = {astro-ph.HE},
       adsurl = {https://ui.adsabs.harvard.edu/abs/2025A&A...696A..36D}
}

@ARTICLE{1971NPhS..229..177P,
       author = {{Penrose}, R. and {Floyd}, R.~M.},
        title = "{Extraction of Rotational Energy from a Black Hole}",
      journal = {Nature Physical Science},
         year = 1971,
        month = feb,
       volume = {229},
       number = {6},
        pages = {177-179},
          doi = {10.1038/physci229177a0},
       adsurl = {https://ui.adsabs.harvard.edu/abs/1971NPhS..229..177P}
}

@ARTICLE{2025A&A...696A..10A,
       author = {{Antonopoulou}, Eleni and {Loules}, Argyrios and {Nathanail}, Antonios},
        title = "{Magnetically arrested disk flux eruption events to describe SgrA* flares}",
      journal = {\aap},
         year = 2025,
        month = apr,
       volume = {696},
          eid = {A10},
        pages = {A10},
          doi = {10.1051/0004-6361/202453456},
archivePrefix = {arXiv},
       eprint = {2501.07521},
 primaryClass = {astro-ph.HE},
       adsurl = {https://ui.adsabs.harvard.edu/abs/2025A&A...696A..10A}
}

@ARTICLE{2024MNRAS.531.3136L,
       author = {{Lin}, Xi and {Yuan}, Feng},
        title = "{Revisiting flares in Sagittarius A* based on general relativistic magnetohydrodynamic numerical simulations of black hole accretion}",
      journal = {\mnras},
         year = 2024,
        month = jul,
       volume = {531},
       number = {3},
        pages = {3136-3150},
          doi = {10.1093/mnras/stae1357},
archivePrefix = {arXiv},
       eprint = {2405.17408},
 primaryClass = {astro-ph.HE},
       adsurl = {https://ui.adsabs.harvard.edu/abs/2024MNRAS.531.3136L}
}

@ARTICLE{2024ApJ...964L..25E,
       author = {{Event Horizon Telescope Collaboration} and {Akiyama}, Kazunori and {Alberdi}, Antxon and {Alef}, Walter and {Algaba}, Juan Carlos and {Anantua}, Richard and {Asada}, Keiichi and {Azulay}, Rebecca and {Bach}, Uwe and {Baczko}, Anne-Kathrin and {Ball}, David and {Balokovic}, Mislav and {Bandyopadhyay}, Bidisha and {Barrett}, John and {Baub{\"o}ck}, Michi and {Benson}, Bradford A. and {Bintley}, Dan and {Blackburn}, Lindy and {Blundell}, Raymond and {Bouman}, Katherine L. and {Bower}, Geoffrey C. and {Boyce}, Hope and {Bremer}, Michael and {Brinkerink}, Christiaan D. and {Brissenden}, Roger and {Britzen}, Silke and {Broderick}, Avery E. and {Broguiere}, Dominique and {Bronzwaer}, Thomas and {Bustamante}, Sandra and {Byun}, Do-Young and {Carlstrom}, John E. and {Ceccobello}, Chiara and {Chael}, Andrew and {Chan}, Chi-kwan and {Chang}, Dominic O. and {Chatterjee}, Koushik and {Chatterjee}, Shami and {Chen}, Ming-Tang and {Chen}, Yongjun and {Cheng}, Xiaopeng and {Cho}, Ilje and {Christian}, Pierre and {Conroy}, Nicholas S. and {Conway}, John E. and {Cordes}, James M. and {Crawford}, Thomas M. and {Crew}, Geoffrey B. and {Cruz-Osorio}, Alejandro and {Cui}, Yuzhu and {Dahale}, Rohan and {Davelaar}, Jordy and {De Laurentis}, Mariafelicia and {Deane}, Roger and {Dempsey}, Jessica and {Desvignes}, Gregory and {Dexter}, Jason and {Dhruv}, Vedant and {Dihingia}, Indu K. and {Doeleman}, Sheperd S. and {Dougal}, Sean Taylor and {Dzib}, Sergio A. and {Eatough}, Ralph P. and {Emami}, Razieh and {Falcke}, Heino and {Farah}, Joseph and {Fish}, Vincent L. and {Fomalont}, Edward and {Ford}, H. Alyson and {Foschi}, Marianna and {Fraga-Encinas}, Raquel and {Freeman}, William T. and {Friberg}, Per and {Fromm}, Christian M. and {Fuentes}, Antonio and {Galison}, Peter and {Gammie}, Charles F. and {Garc{\'\i}a}, Roberto and {Gentaz}, Olivier and {Georgiev}, Boris and {Goddi}, Ciriaco and {Gold}, Roman and {G{\'o}mez-Ruiz}, Arturo I. and {G{\'o}mez}, Jos{\'e} L. and {Gu}, Minfeng and {Gurwell}, Mark and {Hada}, Kazuhiro and {Haggard}, Daryl and {Haworth}, Kari and {Hecht}, Michael H. and {Hesper}, Ronald and {Heumann}, Dirk and {Ho}, Luis C. and {Ho}, Paul and {Honma}, Mareki and {Huang}, Chih-Wei L. and {Huang}, Lei and {Hughes}, David H. and {Ikeda}, Shiro and {Impellizzeri}, C.~M. Violette and {Inoue}, Makoto and {Issaoun}, Sara and {James}, David J. and {Jannuzi}, Buell T. and {Janssen}, Michael and {Jeter}, Britton and {Jiang}, Wu and {Jim{\'e}nez-Rosales}, Alejandra and {Johnson}, Michael D. and {Jorstad}, Svetlana and {Joshi}, Abhishek V. and {Jung}, Taehyun and {Karami}, Mansour and {Karuppusamy}, Ramesh and {Kawashima}, Tomohisa and {Keating}, Garrett K. and {Kettenis}, Mark and {Kim}, Dong-Jin and {Kim}, Jae-Young and {Kim}, Jongsoo and {Kim}, Junhan and {Kino}, Motoki and {Koay}, Jun Yi and {Kocherlakota}, Prashant and {Kofuji}, Yutaro and {Koch}, Patrick M. and {Koyama}, Shoko and {Kramer}, Carsten and {Kramer}, Joana A. and {Kramer}, Michael and {Krichbaum}, Thomas P. and {Kuo}, Cheng-Yu and {La Bella}, Noemi and {Lauer}, Tod R. and {Lee}, Daeyoung and {Lee}, Sang-Sung and {Leung}, Po Kin and {Levis}, Aviad and {Li}, Zhiyuan and {Lico}, Rocco and {Lindahl}, Greg and {Lindqvist}, Michael and {Lisakov}, Mikhail and {Liu}, Jun and {Liu}, Kuo and {Liuzzo}, Elisabetta and {Lo}, Wen-Ping and {Lobanov}, Andrei P. and {Loinard}, Laurent and {Lonsdale}, Colin J. and {Lowitz}, Amy E. and {Lu}, Ru-Sen and {MacDonald}, Nicholas R. and {Mao}, Jirong and {Marchili}, Nicola and {Markoff}, Sera and {Marrone}, Daniel P. and {Marscher}, Alan P. and {Mart{\'\i}-Vidal}, Iv{\'a}n and {Matsushita}, Satoki and {Matthews}, Lynn D. and {Medeiros}, Lia and {Menten}, Karl M. and {Michalik}, Daniel and {Mizuno}, Izumi and {Mizuno}, Yosuke and {Moran}, James M. and {Moriyama}, Kotaro and {Moscibrodzka}, Monika and {Mulaudzi}, Wanga and {M{\"u}ller}, Cornelia and {M{\"u}ller}, Hendrik and {Mus}, Alejandro and {Musoke}, Gibwa and {Myserlis}, Ioannis and {Nadolski}, Andrew and {Nagai}, Hiroshi and {Nagar}, Neil M. and {Nakamura}, Masanori and {Narayanan}, Gopal and {Natarajan}, Iniyan and {Nathanail}, Antonios and {Fuentes}, Santiago Navarro and {Neilsen}, Joey and {Neri}, Roberto and {Ni}, Chunchong and {Noutsos}, Aristeidis and {Nowak}, Michael A. and {Oh}, Junghwan and {Okino}, Hiroki and {Olivares}, H{\`e}ctor and {Ortiz-Le{\'o}n}, Gisela N. and {Oyama}, Tomoaki and {{\"O}zel}, Feryal and {Palumbo}, Daniel C.~M. and {Paraschos}, Georgios Filippos and {Park}, Jongho and {Parsons}, Harriet and {Patel}, Nimesh and {Pen}, Ue-Li},
        title = "{First Sagittarius A* Event Horizon Telescope Results. VII. Polarization of the Ring}",
      journal = {\apjl},
         year = 2024,
        month = apr,
       volume = {964},
       number = {2},
          eid = {L25},
        pages = {L25},
          doi = {10.3847/2041-8213/ad2df0},
       adsurl = {https://ui.adsabs.harvard.edu/abs/2024ApJ...964L..25E}
}

@ARTICLE{2023A&A...677A..67E,
       author = {{El Mellah}, I. and {Cerutti}, B. and {Crinquand}, B.},
        title = "{Reconnection-driven flares in 3D black hole magnetospheres. A scenario for hot spots around Sagittarius A*}",
      journal = {\aap},
         year = 2023,
        month = sep,
       volume = {677},
          eid = {A67},
        pages = {A67},
          doi = {10.1051/0004-6361/202346781},
archivePrefix = {arXiv},
       eprint = {2305.01689},
 primaryClass = {astro-ph.HE},
       adsurl = {https://ui.adsabs.harvard.edu/abs/2023A&A...677A..67E}
}

@ARTICLE{2015PhPl...22a2902K,
       author = {{Kumar}, Dinesh and {Bhattacharyya}, R. and {Smolarkiewicz}, P.~K.},
        title = "{Repetitive formation and decay of current sheets in magnetic loops: An origin of diverse magnetic structures}",
      journal = {Physics of Plasmas},
         year = 2015,
        month = jan,
       volume = {22},
       number = {1},
          eid = {012902},
        pages = {012902},
          doi = {10.1063/1.4905643},
       adsurl = {https://ui.adsabs.harvard.edu/abs/2015PhPl...22a2902K}
}
\bibliographystyle{bibtex/aa}

\begin{appendix}

\section{Radiative cooling terms}\label{appenC}
\renewcommand{\thefigure}{A\arabic{figure}}

The total cooling rate for an optically thin gas is determined as:
 \begin{eqnarray}
 q^- = q^-_{\text{br}} + \eta_Cq^-_{\text{cs}},
 \end{eqnarray}
 where $\eta_C$ denotes the Compton enhancement factor, which is explicitly given as \cite{1995ApJ...452..710N}:
\begin{eqnarray}
\eta(\nu_c) = 1 + \frac{P(A - 1)}{1 - P A} \left( 1 - \left( \frac{3k_B T_e}{h \nu_c} \right)^{\eta_1} \right),
 \end{eqnarray}
 the terms in the Compton enhancement factor are defined as follows: $
 P = 1 - \exp(-\tau_{\text{es}}), \quad A = 1 + 4\Theta_e + 16\Theta_e^2, \quad \eta_1 = 1 + \ln P / \ln A,
 \tau_{\text{es}} = 2\sigma_T n_e H
 $. Where $\sigma_T$ is the Thomson cross-section of the electron, $n_e$ is the electron number density, $H$ is the local scale height.
 
To compute the bremsstrahlung cooling rate, we adopt the formulations presented by \cite{1996ApJ...465..312E}. Specifically, the free-free bremsstrahlung cooling rate for an ionized plasma, consisting of electrons and ions, is expressed as:
\begin{eqnarray}
q^-_{\text{br}} = q_{\text{br}}^{ei} + q_{\text{br}}^{ee},
\end{eqnarray}
where the explicit forms of the individual components are given by:
{\small
\begin{eqnarray}
&q_{\text{br}}^{ei} = 1.48 \times 10^{-22} n_i n_e \times\nonumber \\
&\left\{
\begin{array}{ll}
\displaystyle\frac{4\sqrt{2\Theta_e}}{\pi^{3/2}} 
\left(1 + 1.781 \Theta_e^{1.34}\right), & \Theta_e < 1 \\[6pt]
\displaystyle\frac{9\Theta_e}{2\pi} 
\left[\ln(1.123\Theta_e) + 0.48\right] + 1.5, & \text{otherwise}
\end{array}
\right.
\end{eqnarray}
\begin{eqnarray}
q_{\text{br}}^{ee} =
\left\{
\begin{array}{ll}
\displaystyle 2.56 \times 10^{-22} \, n_e^2 \Theta_e^{3/2} 
\left(1 + 1.1\Theta_e + \Theta_e^2 - 1.25\Theta_e^{5/2} \right), & \Theta_e < 1 \\[6pt]
\displaystyle 3.24 \times 10^{-22} \, n_e^2 \Theta_e 
\left[\ln(1.123\Theta_e) + 1.28\right], & \text{otherwise}
\end{array}
\right.
\end{eqnarray}
}
Here, $\Theta_e := k_B T_e / m_e c^2$ is the dimensionless electron temperature.

In the presence of a strong magnetic field, the hot electrons in the accretion flow radiate via the thermal synchrotron process. The rate of synchrotron emission, as formulated by \cite{1996ApJ...465..312E}, is expressed as:
\begin{eqnarray}
q^-_{\text{cs}} &=& \frac{2\pi k_B T_i \nu_c^3}{3 H c^2} + 6.76 \times 10^{-28} \frac{n_e}{K_2(1/\Theta_e) a_1^{1/6}}\times \nonumber \label{eq11} \\
&& \left[\frac{1}{a_4^{11/2}} \Gamma\left(\frac{11}{2}, a_4 \nu_c^{1/3}\right)
	+ \frac{a_2}{a_4^{19/4}} \Gamma\left(\frac{19}{4}, a_4 \nu_c^{1/3}\right) \right.  \\
&& \left. + \frac{a_3}{a_4^4} \left(a_3^4 \nu_c + 3a_2^2 \nu_c^{2/3} + 6a_4 \nu_c^{1/3} + 6\right) e^{-a_4 \nu_c^{1/3}} \right] \nonumber
\end{eqnarray}
where $H$ is the local scale height, estimated from the gradient of the electron temperature as $H = T_e^4$ / $|\nabla T_e^4|$. The function $K_2$ represents the modified Bessel function of the second kind. The coefficients $a_1$ through $a_4$ in Eq.~\ref{eq11} are as defined in \cite{1996ApJ...465..312E}.

\section{Resolution test}\label{appenD}
\renewcommand{\thefigure}{B\arabic{figure}}

\begin{figure}
    \centering
    \includegraphics[width=\linewidth]{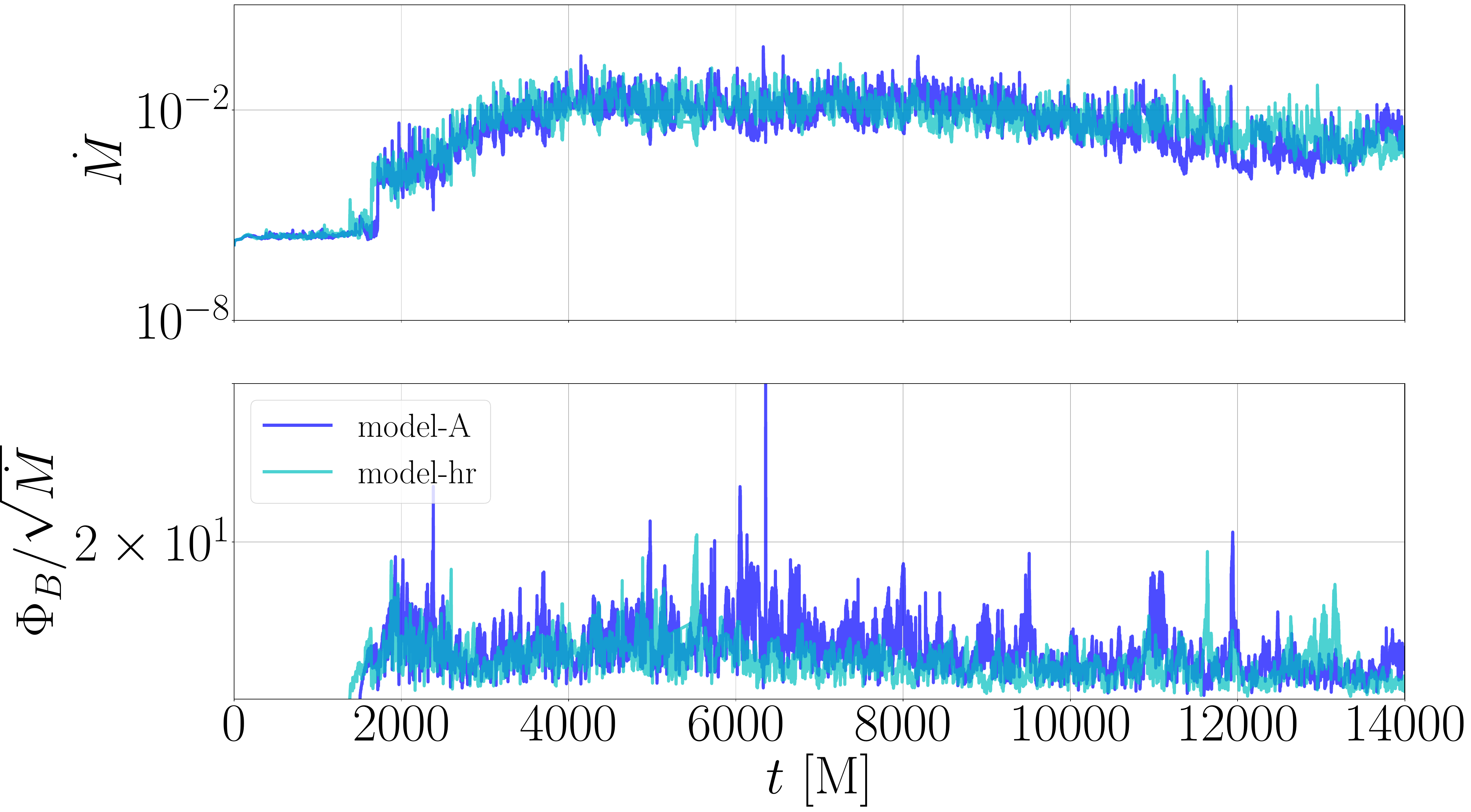}
    \caption{Time evolution of the mass accretion rate (top panel) and the magnetic flux (bottom panel) for model A and the high-resolution model (hr).}
    \label{fig:Mdot_hr}
\end{figure}

To investigate the effects of numerical resolution, we conduct two-dimensional GRMHD simulations of magnetized accretion flows onto a rotating BH, incorporating radiative cooling, at different grid resolutions.  Model $\tt{hr}$ has the same initial condition as model A, differing only in resolution. We adopt a grid size of $(N_r, N_\theta) = 1024 \times 512$ as the standard resolution and We adopted five levels of static mesh refinement with a base resolution of $(N_r, N_\theta) = 256 \times 128$. The second refinement level is applied within the radial range $100 < r/r_g < 200$, the third within $90 < r/r_g < 100$, the fourth within $60 < r/r_g < 90$, and the fifth within $3 < r/r_g < 60$. As a result, the effective resolution in the regions of prime interest reaches $(N_r, N_\theta) = 4096 \times 2048$, allowing us to capture the dynamics near the current sheet with significantly enhanced spatial resolution. As shown in Fig.~\ref{fig:Mdot_hr}, the evolutionary trends between different resolution cases show excellent agreement. 

\begin{figure*}
    \centering
    \includegraphics[width=0.9\linewidth]{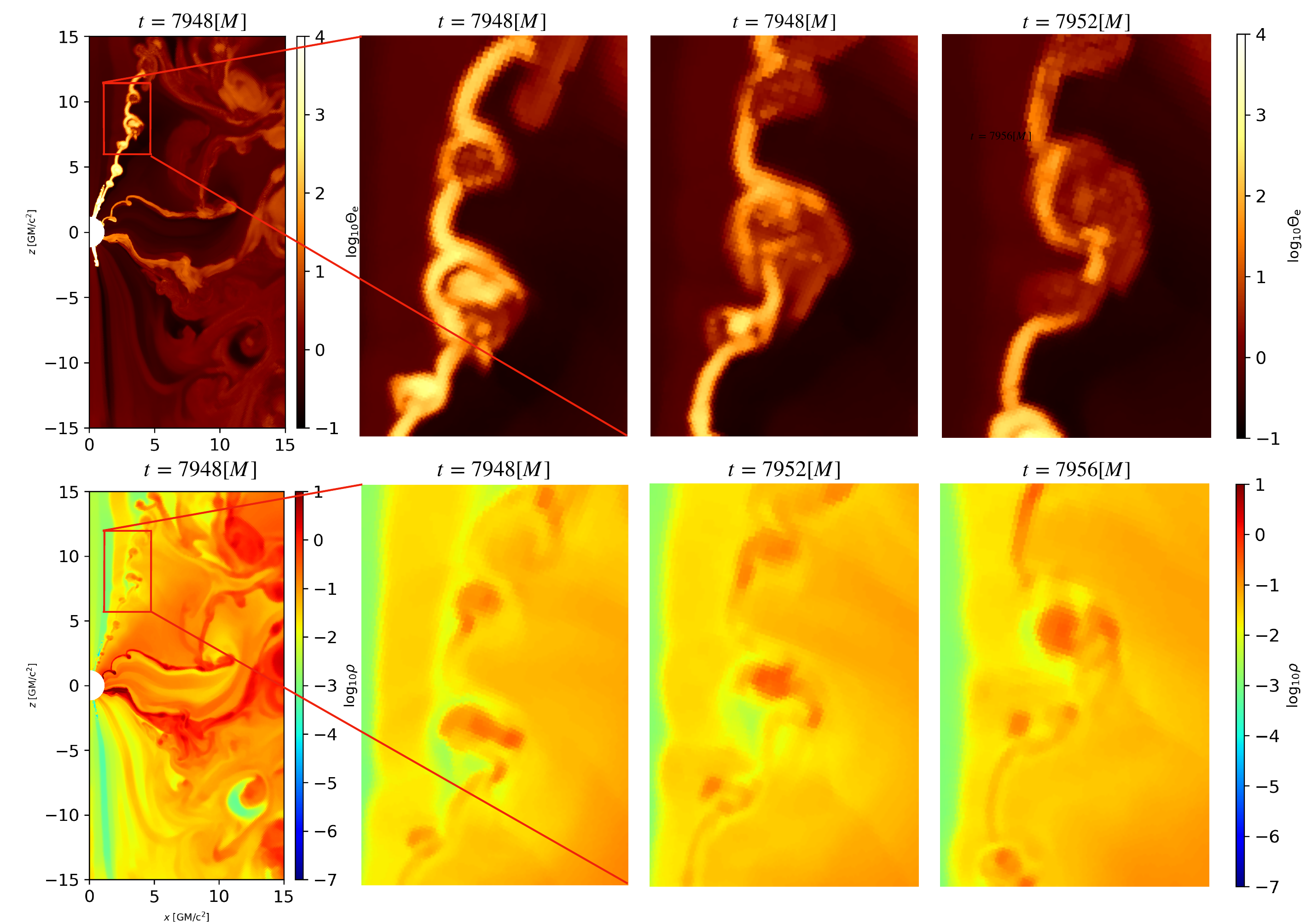}
    \caption{The snapshots show the distribution of electron temperature and density at standard resolution.}
    \label{fig:lr}
\end{figure*}

In the meanwhile, we also observed radiative collapse induced by radiative cooling, shown in Fig.~\ref{fig:lr}. This confirms that radiative collapse is a general and resolution-insensitive feature of the system, rather than a numerical artifact.

\end{appendix}

\end{document}